%% file: paper.tex
\documentclass{JHEP}    
\usepackage{epsfig}

\input texpref.tex

\newcommand{\be}{\begin{equation}}
\newcommand{\ee}{\end{equation}}
\newcommand{\bea}{\begin{eqnarray}}
\newcommand{\eea}{\end{eqnarray}}
\newcommand{\bean}{\begin{eqnarray*}}
\newcommand{\eean}{\end{eqnarray*}}

\preprint{MIT-CTP-3184\\ \\ {\tt hep-th/}}

\title{Toric Duality as Seiberg Duality and Brane Diamonds}
\author{
Bo Feng$^1$, Amihay Hanany$^1$, Yang-Hui He$^1$ and Angel M. Uranga$^2$
\footnote{
Research supported in part
by the CTP and the LNS of MIT and the U.S. Department of Energy
under cooperative research agreement \# DE-FC02-94ER40818.
A.~H. is also supported by an A. P. Sloan Foundation Fellowship, the
Reed Fund and a DOE OJI award. Y.-H.~H. is also supported by the Presidential
Fellowship of M.I.T. A.~M.~U. is supported by the TH Division, CERN}
\\
~\\
$^1$ Center for Theoretical Physics,
\\ Massachusetts Institute of Technology,\\
Cambridge, MA 02139, USA.\\
~\\
$^2$ TH Division, CERN \\ CH-1211 Geneva 23, Switzerland\\
~\\
\email{fengb, hanany, yhe@ctp.mit.edu, Angel.Uranga@cern.ch}
}

\abstract{We use field theory and brane diamond techniques to
demonstrate that Toric Duality is Seiberg duality for ${\cal N}=1$
theories with toric moduli spaces.
This resolves the puzzle concerning the 
physical meaning of
Toric Duality as proposed in our earlier work.
Furthermore, using this strong connection we arrive at
three new phases which can not be thus far obtained by the so-called
``Inverse Algorithm'' applied to partial resolution of
$\C^3/(\Z_3 \times \Z_3)$. The standing proposals of Seiberg duality as 
diamond duality in the work by Aganagic-Karch-L\"ust-Miemiec
are strongly supported and
new diamond configurations for these
singularities are obtained as a byproduct.
We also make some remarks about the relationships between
Seiberg duality and Picard-Lefschetz monodromy.
}
\begin{document}
\section{Introduction}
Witten's gauge linear sigma approach \cite{Witten} to ${\cal N}=2$
super-conformal theories has provided deep insight not only to the
study of the phases of the field theory but also to the understanding
of the mathematics of Geometric Invariant Theory 
quotients in toric geometry. Thereafter, the
method was readily applied to the study of the ${\cal N}=1$ supersymmetric 
gauge theories on D-branes at singularities 
\cite{DGM,Morrison,Uranga,Chris}.
Indeed the classical moduli space of the gauge theory corresponds
precisely to the spacetime which the D-brane probes transversely. In 
light of this therefore, toric geometry has been widely used in the study 
of the moduli space of vacua of the gauge theory living on D-brane probes.

The method of encoding the gauge theory data into the moduli
data, or more specifically, the F-term and D-term information into the
toric diagram of the algebraic variety describing the moduli space,
has been well-established \cite{DGM,Morrison}. The reverse, of
determining the SUSY gauge theory data in terms of a given toric
singularity upon which the D-brane probes, has also been addressed
using the method partial resolutions of abelian quotient
singularities. Namely, a general non-orbifold singularity is regarded as 
a partial resolution of a worse, but orbifold, singularity. This ``Inverse 
Procedure'' was formalised into a linear optimisation algorithm, easily 
implementable on computer, by \cite{toric}, and was subsequently checked 
extensively in \cite{Sarkar}.

One feature of the Inverse Algorithm is its non-uniqueness, viz., that
for a given toric singularity, one could in theory construct countless
gauge theories. This means that there are classes of gauge theories which 
have identical toric moduli space in the IR. Such a salient feature was 
dubbed in \cite{toric} as {\bf toric duality}. Indeed in a follow-up work, 
\cite{phases} attempted to analyse this duality in detail, concentrating 
in particular on a method of fabricating dual theories which are physical, 
in the sense that they can be realised as world-volume theories on 
D-branes. Henceforth, we shall adhere to this more restricted meaning of 
toric duality.

Because the details of this method will be clear in later examples we 
shall not delve into the specifics here, nor shall we devote too much 
space reviewing the algorithm. Let us highlight the key points. The gauge 
theory data of D-branes probing Abelian orbifolds is well-known (see e.g. 
the appendix of \cite{phases}); also any toric diagram can be embedded
into that of such an orbifold (in particular any toric local Calabi-Yau 
threefold $D$ can be embedded into $\IC^3/(\IZ_n \times \IZ_n)$ for 
sufficiently large $n$. We can then obtain the subsector of orbifold 
theory that corresponds the gauge theory constructed for $D$. This is the 
method of ``Partial Resolution.''

A key point of \cite{phases} was the application of the well-known 
mathematical fact that the toric diagram $D$ of any toric variety has an 
inherent ambiguity in its definition: namely any unimodular transformation 
on the lattice on which $D$ is defined must leave $D$ invariant. In other 
words, for threefolds defined in the standard lattice $\IZ^3$, any 
$SL(3;\IC)$ transformation on the vector endpoints of the defining toric 
diagram gives the same toric variety. Their embedding into the diagram of 
a fixed Abelian orbifold on the other hand, certainly is different. Ergo, 
the gauge theory data one obtains in general are vastly different, even 
though per constructio, they have the same toric moduli space.

What then is this ``toric duality''? How clearly it is defined
mathematically and yet how illusive it is as a physical phenomenon.
The purpose of the present writing is to make the first leap toward
answering this question. In particular, we shall show, using brane
setups, and especially brane diamonds, that known cases for toric
duality are actually interesting realisations of Seiberg Duality.
Therefore the mathematical equivalence of moduli spaces for different 
quiver gauge theories is related to a real physical equivalence of the 
gauge theories in the far infrared.

The paper is organised as follows. In Section 2, we begin with an
illustrative example of two torically dual cases of a generalised
conifold. These are well-known to be Seiberg dual theories as seen
from brane setups. Thereby we are motivated to conjecture in Section 3
that toric duality is Seiberg duality. We proceed to check this
proposal in Section 4 with all the known cases of torically dual
theories and have successfully shown that the phases of the partial
resolutions of $\IC^3/(\IZ_3 \times \IZ_3)$ constructed in
\cite{toric} are indeed Seiberg dual from a field theory analysis.
Then in Section 6 we re-analyse these examples from the perspective of
brane diamond configurations and once again obtain strong support of
the statement. From rules used in the diamond dualisation, 
we extracted a so-called ``quiver duality'' which explicits Seiberg
duality as a transformation on the matter adjacency matrices. Using
these rules we are able to extract more phases of theories not yet
obtained from the Inverse Algorithm. In a more geometrical vein, in
Section 7, we remark the
connection between Seiberg duality and Picard-Lefschetz and point out
cases where the two phenomena may differ. Finally we finish with
conclusions and prospects in Section 8.

While this manuscript is about to be released, we became aware of the
nice work \cite{BP}, which discusses similar issues.
\section{An Illustrative Example}
We begin with an illustrative example that will demonstrate how Seiberg 
Duality is realised as toric duality.
\subsection{The Brane Setup}
The example is the well-known generalized conifold described 
as the hypersurface $xy=z^2w^2$ in $\IC^4$, and which can be obtained 
as a $\IZ_2$ quotient of the famous conifold $xy=zw$ by the action 
$z\to -z, w\to -w$. The gauge theory on the D-brane sitting at such a 
singularity can be established by orbifolding the conifold gauge theory in 
\cite{kw}, as in \cite{conuran}. Also, it can be derived by another method 
alternative to the Inverse Algorithm, namely performing a T-duality to a 
brane setup with NS-branes and D4-branes \cite{conuran,dasmuk}. Therefore 
this theory serves as an excellent check on our methods.

The setup involves stretching D4 branes (spanning 01236) between 2 pairs 
of NS and NS$'$ branes (spanning 012345 and 012389, respectively), with 
$x^6$ parameterizing a circle. These configurations are analogous to 
those in \cite{EGK}.
There are in fact two inequivalent brane setups (a) and (b) (see 
\fref{z2w2brane}), differing in the way the NS- and NS$'$-branes are 
ordered in the circle coordinate.
%
\EPSFIGURE[h]{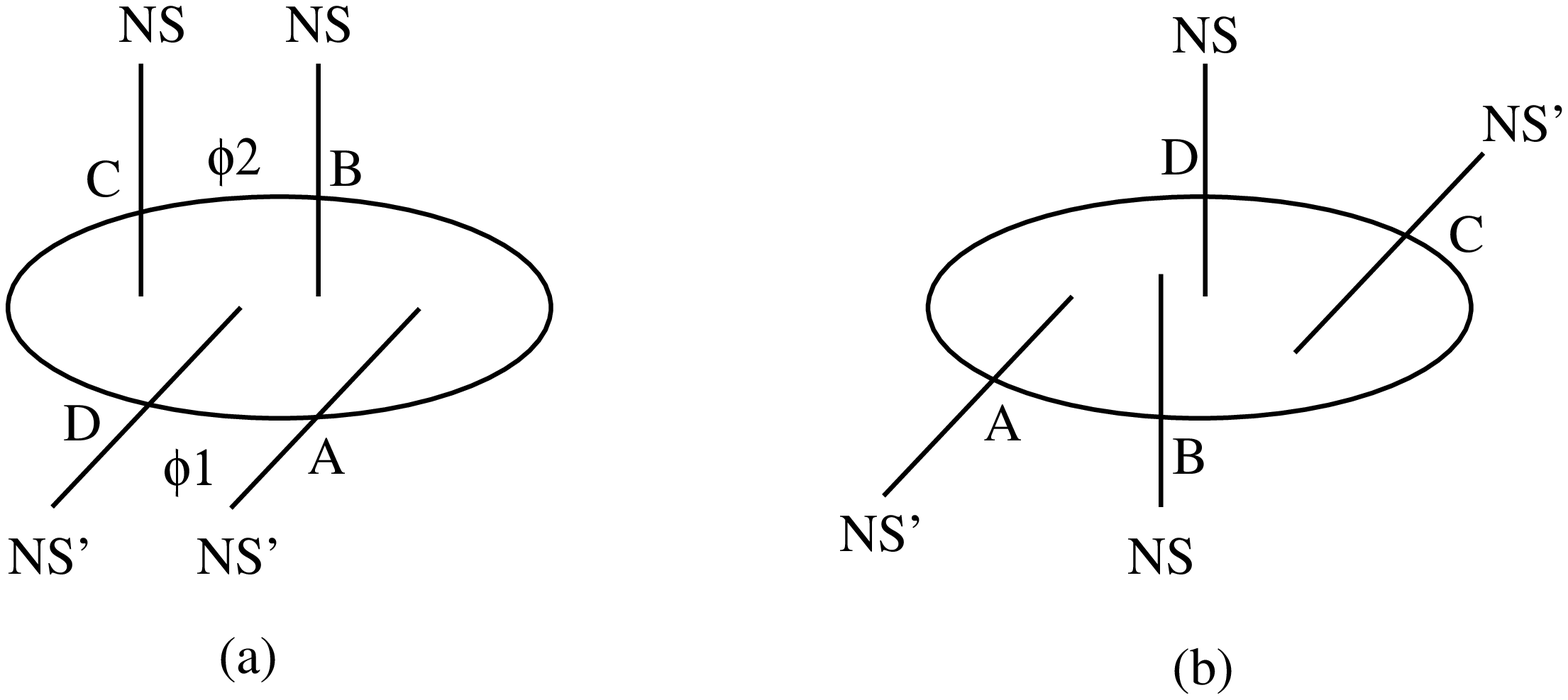,width=14cm}
{The two possible brane setups for the generalized conifold $xy=z^2
w^2$. They are related to each other passing one NS-brane through an
NS'-brane. $A_i,B_i,C_i,D_i$ $i=1,2$ are
bifundamentals while  $\phi_1,\phi_2$ are
two adjoint fields.
\label{z2w2brane}
}
Using standard rules \cite{HW,EGK}, we see from the figure that there are 
4 product gauge groups (in the Abelian case, it is simply $U(1)^4$. As for 
the matter content, theory (a) has 8 bi-fundamental chiral multiplets 
$A_i$, $B_i$, $C_i$, $D_i$ $i=1,2$ (with charge $(+1,-1)$ and $(-1,+1)$ 
with respect to adjacent $U(1)$ factors) and 2 adjoint chiral multiplets
$\phi_{1,2}$ as indicated. On the other hand (b) has only 8 
bi-fundamentals, with charges as above. The superpotentials are 
respectively \cite{ahan,conuran}
\bean
(a)~~~ & W_a= & -A_1 A_2 B_1 B_2 +B_1 B_2 \phi_2-C_1 C_2 \phi_2 
	+C_1 C_2 D_1 D_2 - D_1 D_2  \phi_1 + A_1 A_2 \phi_1,
\\
(b)~~~ & W_b= & A_1 A_2 B_1 B_2-B_1 B_2  C_1 C_2 +  C_1 C_2 D_1 D_2 
	-D_1 D_2 A_1 A_2  
\eean
With some foresight, for comparison with the results later, we rewrite 
them as
\be
\label{orbi-coni-1-1} 
W_a=(B_1 B_2 - C_1 C_2)(\phi_2- A_1 A_2)+ (A_1 A_2 - D_1 D_2)
(\phi_1 -C_1 C_2)
\ee
\be
\label{orbi-coni-2-1}
W_b = (A_1 A_2 -C_1 C_2) (B_1 B_2 -D_1 D_2)
\ee
\subsection{Partial Resolution} 
Let us see whether we can reproduce these field theories with the Inverse 
Algorithm. The toric diagram for $xy=z^2w^2$ is given in the very left of 
\fref{z2w2toric}. Of course, the hypersurface is three complex-dimensional 
so there is actually an undrawn apex for the toric diagram, and each of 
the nodes is in fact a three-vector in $\IZ^3$. Indeed the fact that it is 
locally Calabi-Yau that guarantees all the nodes to be coplanar. The 
next step is the realisation that it can be embedded into the well-known 
toric diagram for the Abelian orbifold $\IC^3/(\IZ_3 \times \IZ_3)$ 
consisting of 10 lattice points. The reader is referred to \cite{toric,phases} 
for the actual co\"ordinates of the points, a detail which, though crucial, 
we shall not belabour here.

The important point is that there are six ways to embed our toric diagram 
into the orbifold one, all related by $SL(3;\IC)$ transformations. This is 
indicated in parts (a)-(f) of \fref{z2w2toric}. We emphasise that these 
six diagrams, drawn in red, are {\em equivalent} descriptions of $xy=z^2w^2$ 
by virtue of their being unimodularly related; therefore they are all 
candidates for toric duality.
\EPSFIGURE[h]{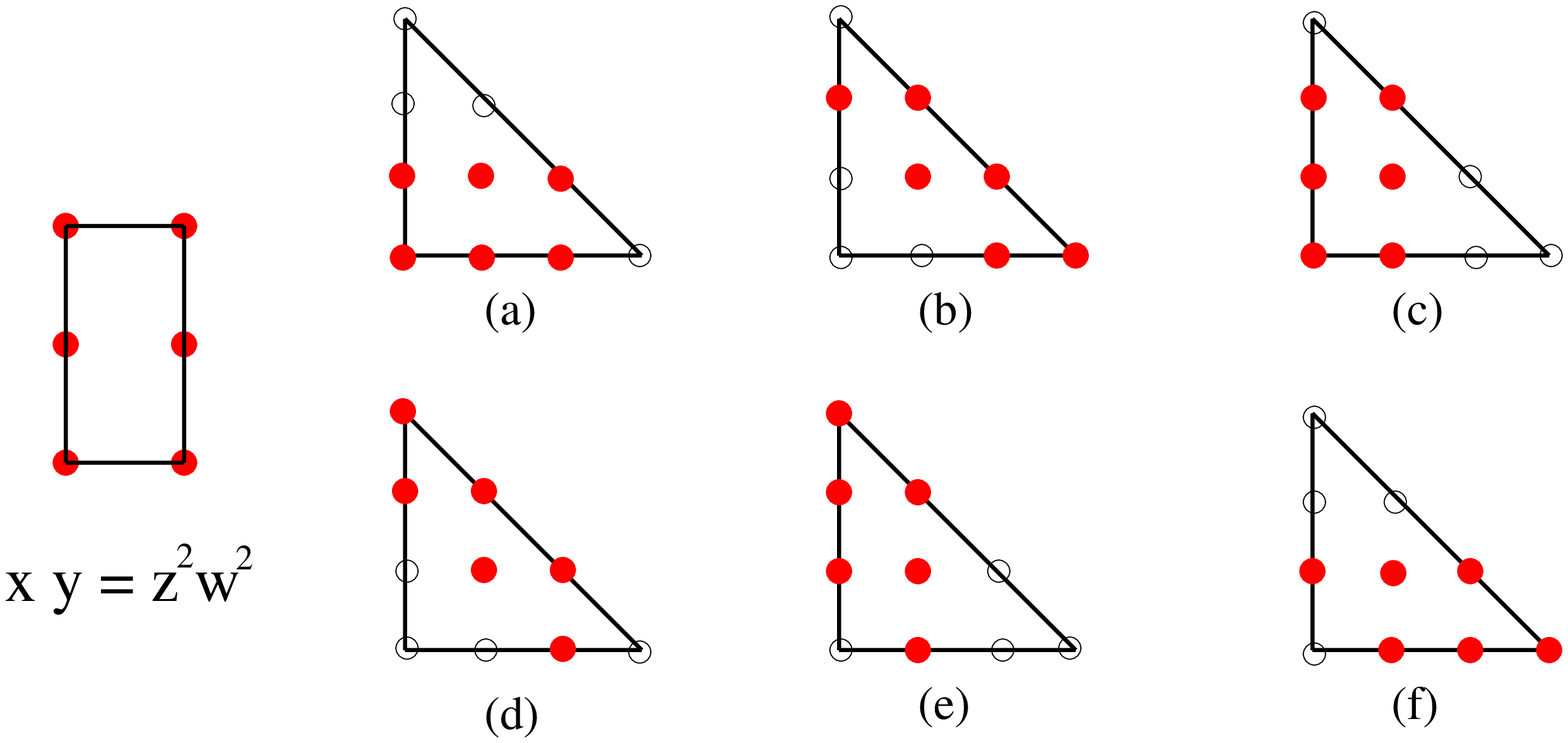,width=14cm}
{The standard toric diagram for the generalized conifold $xy=uv=z^2$
(far left). To the right are six $SL(3;\IC)$ transformations (a)-(f)
thereof (drawn in red) and hence are equivalent toric diagrams for the
variety. We embed these six diagrams into the Abelian orbifold
$\IC^3/(\IZ_3 \times \IZ_3)$ in order to perform partial resolution and
thus the gauge theory data.
\label{z2w2toric}
}

Now we use our 
Inverse Algorithm, by partially resolving $\IC^3/(\IZ_3 \times 
\IZ_3)$, to obtain the gauge theory data for the D-brane probing 
$xy=z^2w^2$. In summary, after exploring the six possible partial 
resolutions, we find that cases (a) and (b) give identical results, while 
(c,d,e,f) give the same result which is inequivalent from (a,b). Therefore 
we conclude that cases (a) and (c) are inequivalent torically dual 
theories for $xy=z^2 w^2$. In the following we detail the data for these 
two contrasting cases. We refer the reader to \cite{toric,phases} for details and 
notation.
\EPSFIGURE[ht]{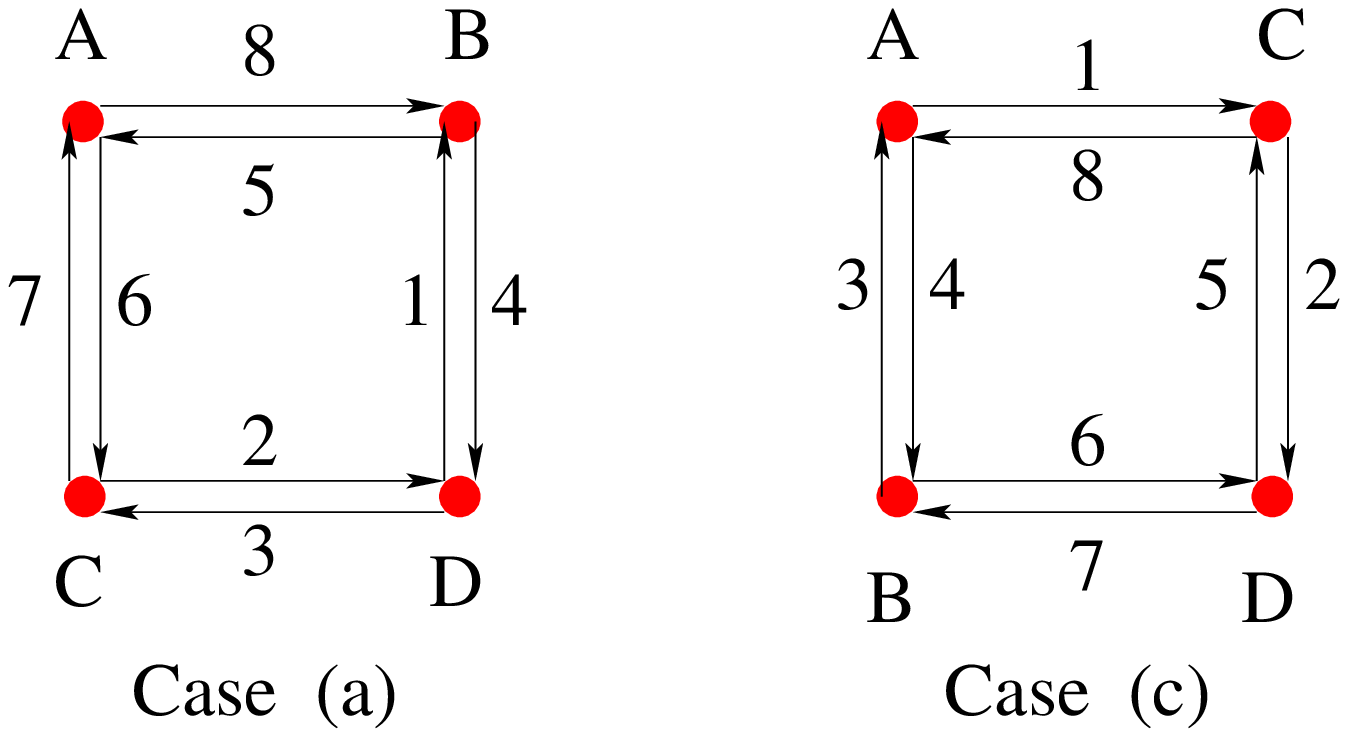,width=10cm}
{The quiver diagram encoding the matter content of Cases (a) and (c)
of \fref{z2w2toric}.
\label{z2w2quiver}
}
\subsection{Case (a) from Partial Resolution}
For case (a), the matter content is encoded the $d$-matrix  
which indicates the charges of the 8 bi-fundamentals under the 4 gauge
groups. This is the incidence matrix for the quiver diagram drawn in part 
(a) of \fref{z2w2quiver}. 
$$
\tmat{
         &  X_1 & X_2 & X_3 & X_4 & X_5 & X_6 & X_7 & X_8  \cr
U(1)_A   &   0  &  0  &  0  &  0  &  1  &  -1 &  1  &  -1  \cr
U(1)_B   &   1  &  0  &  0  &  -1 & -1  &  0  &  0  &   1  \cr
U(1)_C   &   0  & -1  &  1  &  0  &  0  &   1 & -1  &  0   \cr
U(1)_D   &   -1 & 1   &  -1 &  1  &  0  &  0  &  0  &  0   \cr
}
$$
On the other hand, the F-terms are encoded in the $K$-matrix
$$
\tmat{
  X_1 & X_2 & X_3 & X_4 & X_5 & X_6 & X_7 & X_8  \cr
1 &0 &1 &0 &0 &0 &0 &0  \cr
1 &1 &0 &0 &0 &0 &0 &0  \cr
0 &1 &0 &1 &0 &0 &0 &0  \cr
0 &0 &0 &0 &1 &0 &1 &0  \cr
0 &0 &0 &0 &1 &1 &0 &0  \cr
0 &0 &0 &0 &0 &1 &0 &1  \cr
}
$$
From $K$ we get two relations $X_5 X_8 =X_6 X_7 $ and $X_1 X_4=X_2 X_3$ 
(these are the relations one must impose on the quiver to obtain the final 
variety; equivalently, they correspond to the F-term constraints arising 
from the superpotential). Notice that here each term is chargeless under 
all 4 gauge groups, so when we integrate back to get the superpotential,
we should multiply by chargeless quantities also\footnote{In more general
situations the left- and right-hand sides may not be singlets, but transform 
in the same gauge representation.}.

The relations must come from the F-flatness $\frac{\partial}{\partial X_i} 
W = 0$ and thus we can use these relations to integrate back to the 
superpotential $W$. However we meet some ambiguities
\footnote{The ambiguities arise because in the abelian case (toric 
language) the adjoints are chargeless. In fact, no ambiguity arises if 
one performs the Higgsing associated to the partial resolution in the 
non-abelian case. We have performed this exercise in cases (a) and (c), 
and verified the result obtained by the different argument offered in the 
text.}. In principle we can have two 
different choices:
\bean
(i)~~~~ & W_1= & (X_5 X_8 -X_6 X_7)(X_1 X_4-X_2 X_3)  \\
(ii)~~~~~& W_2= & \psi_1 (X_5 X_8 -X_6 X_7) +\psi_2 (X_1 X_4-X_2 X_3)
\eean
where for now $\psi_i$ are simply chargeless fields.

We shall evoke physical arguments to determine which is correct. Expanding 
(i) gives $W_1=X_5 X_8 X_1 X_4-X_6 X_7 X_1 X_4-X_5 X_8 X_2X_3+ 
X_6 X_7 X_2 X_3$. Notice the term $X_6 X_7 X_1 X_4$: there is no common 
gauge group under which there four fields are charged, i.e. these 4 
arrows (q.~v. \fref{z2w2quiver}) do not intersect at a single node. This 
makes (i) very unnatural and exclude it.

Case (ii) does not have the above problem and indeed all four fields 
$X_5 ,X_8 ,X_6, X_7$ are charged under the $U(1)_A$ gauge group, so 
considering $\psi_1$ to be an adjoint of $U(1)_A$, we do obtain a physically 
meaningful interaction. Similarly $\psi_2$ will be the adjoint of $U(1)_D$, 
interacting with $X_1 ,X_4 ,X_2, X_3$.

However, we are not finish yet. From \fref{z2w2quiver} we see that 
$X_5, X_8,X_1, X_4$ are all charged under $U(1)_B$, while $X_6, X_7,X_2 ,X_3$ 
are all charged under $U(1)_C$. From a physical point of view, there 
should be some interaction terms between these fields. Possibilities are
$X_5 X_8 X_1 X_4$ and $X_6 X_7 X_2 X_3$. To add these terms into $W_2$ is 
very easy, we simply perform the following replacement:\footnote{Here we  
choose the sign purposefully for later convenience. However, we do need, 
for the cancellation of the unnatural interaction term $X_1 X_4 X_6 X_7$, 
that they both have the same sign.} $\psi_1\longrightarrow \psi_1-X_1 
X_4,~~~~~~\psi_2\longrightarrow \psi_2-X_6 X_7.$ Putting everything 
together, we finally obtain that Case (a) has matter content as described 
in \fref{z2w2quiver} and the superpotential
\beq
\label{z2w2casea}
W=(\psi_1-X_1 X_4)(X_5 X_8 -X_6 X_7) +(\psi_2-X_6 X_7)(X_1 X_4-X_2 X_3)
\eeq

This is precisely the theory (a) from the brane setup in the last
section! Comparing \eref{z2w2casea} with (\ref{orbi-coni-1-1}), we see
that they are exact same under the following redefinition of variables:
$$
\ba{ccccc}
B_1,B_2  \Longleftrightarrow   X_5, X_8  &\qquad &
C_1,C_2  \Longleftrightarrow   X_6,  X_7  &\qquad &
D_1,D_2  \Longleftrightarrow   X_2, X_3  \cr
A_1,A_2  \Longleftrightarrow   X_1,  X_4 &\qquad &
\phi_2 \Longleftrightarrow   \psi_1  & \qquad&
\phi_1 \Longleftrightarrow   \psi_2  \cr 
\ea
$$

In conclusion, case (a) of our Inverse Algorithm reproduces the
results of case (a) of the brane setup.
\subsection{Case (c) from Partial Resolution}
For case (c), the matter content is given by the quiver in
\fref{z2w2quiver}, which has the charge matrix $d$ equal to
$$
\tmat{
         &  X_1 & X_2 & X_3 & X_4 & X_5 & X_6 & X_7 & X_8  \cr
U(1)_A   &   -1 &  0  & -1  &  1  &  0  &  0  &  0  &  1   \cr
U(1)_B   &   0  &  0  &  1  &  -1 &  0  & -1  &  1  &  0   \cr
U(1)_C   &   1  & -1  &  0  &  0  &  1  &  0  &  0  & -1   \cr
U(1)_D   &   0  & 1   &  0  &  0  & -1  &  1  & -1  &  0   \cr
}
$$
This is precisely the matter content of case (b) of the brane setup.
The F-terms are given by
$$
K=
\tmat{
  X_1 & X_2 & X_3 & X_4 & X_5 & X_6 & X_7 & X_8  \cr
0 &1  &0  &1  &0  &0  &0  &0  \cr
1 &0  &0  &0  &0  &0  &1  &0  \cr
1 &0  &0  &0  &0  &1  &0  &0  \cr
0 &1  &1  &0  &0  &0  &0  &0  \cr
0 &0  &1  &0  &1  &0  &0  &0  \cr
0 &0  &0  &0  &0  &1  &0  &1  \cr
}
$$
From it we can read out the relations $X_1 X_8= X_6 X_7$ and $X_2 X_5 = 
X_3 X_4$. Again there are two ways to write down the superpotential
\bean
(i)~~~~ & W_1= & (X_1 X_8 -X_6 X_7)(X_3 X_4-X_2 X_5)  \\
(ii)~~~~~& W_2= & \psi_1 (X_1 X_8 -X_6 X_7) +\psi_2 (X_3 X_4-X_2 X_5)
\eean
In this case, because $X_1$, $X_8$, $X_6$, $X_7$ are not charged under
any common gauge group, it is impossible to include any adjoint field 
$\psi$ to give a physically meaningful interaction and so (ii) is unnatural. We are left the
superpotential $W_1$. Indeed, comparing with (\ref{orbi-coni-2-1}),
we see they are identical under the redefinitions
$$
\ba{ccc}
A_1,A_2 \Longleftrightarrow X_1, X_8 &\qquad &
B_1,B_2 \Longleftrightarrow X_3, X_4 \\
C_1,C_2 \Longleftrightarrow X_6, X_7 &\qquad &
D_1,D_2 \Longleftrightarrow X_2, X_5 \\
\ea
$$
Therefore we have reproduced case (b) of the brane setup.

\medskip

What have we achieved? We have shown that toric duality due to inequivalent 
embeddings of unimodularly related toric diagrams for the generalized 
conifold $xy = z^2w^2$ gives two inequivalent physical world-volume 
theories on the D-brane probe, exemplified by cases (a) and (c). On the 
other hand, there are two T-dual brane setups for this singularity, also 
giving two inequivalent field theories (a) and (b). Upon comparison, case (a)
(resp. (c)) from the Inverse Algorithm beautifully corresponds to case (a) 
(resp. (b)) from the brane setup. Somehow, a seemingly harmless trick in 
mathematics relates inequivalent brane setups. In fact we can say much 
more.
\section{Seiberg Duality versus Toric Duality}
As follows from \cite{EGK}, the two theories from the brane setups are 
actually related by Seiberg Duality \cite{Seiberg}, as pointed out in 
\cite{conuran} (see also \cite{Unge,uranquiver}. Let us first review the 
main features of this famous duality, for unitary gauge groups.

Seiberg duality is a non-trivial infrared equivalence of ${\cal 
N}=1$ supersymmetric field theories, which are different in the 
ultraviolet, but flow the the same interacting fixed point in the 
infrared. In particular, the very low energy features of the different 
theories, like their moduli space, chiral ring, global symmetries, agree 
for Seiberg dual theories. Given that toric dual theories, by definition, 
have identical moduli spaces, etc , it is natural to propose a connection 
between both phenomena.

The prototypical example of Seiberg duality is ${\cal N}=1$ $SU(N_c)$ 
gauge theory with $N_f$ vector-like fundamental flavours, and no 
superpotential. The global chiral symmetry is $SU(N_f)_L\times SU(N_f)_R$, 
so the matter content quantum numbers are 
\[
\ba{c|ccc}
	& SU(N_c)  &  SU(N_f)_L  &  SU(N_f)_R \\ \hline
Q   	& \fund  &   \fund   &      1 \\
Q'	& \antifund &    1     &     \antifund \\
\ea
\]
In the conformal window, $3N_c/2\leq N_f\leq 3N_c$, the theory flows to an 
interacting infrared fixed point. The dual theory, flowing to the same 
fixed point is given ${\cal} N=1$ $SU(N_f-N_c)$ gauge theory with $N_f$ 
fundamental flavours, namely
\[
\ba{c|ccc}
     	&SU(N_f-N_c)   	&SU(N_f)_L   	&SU(N_f)_R \\ \hline
q      	& \fund       	&\antifund	&1 \\
q'     	& \antifund	&1        	&\fund \\
M       &1         	&\fund     	&\antifund \\
\ea
\]
and superpotential $W = M q q'$. From the matching of chiral rings, the 
`mesons' $M$ can be thought of as composites $QQ'$ of the original quarks.

\medskip

It is well established \cite{EGK}, that in an ${\cal N}=1$ (IIA) brane
setup for the four dimensional theory such as \fref{z2w2brane}, Seiberg 
duality is realised as the crossing of 2 non-parallel NS-NS$'$ branes. In 
other words, as pointed out in \cite{conuran}, cases (a) and (b) are in 
fact a Seiberg dual pair. Therefore it seems that the results from the 
previous section suggest that toric duality is a guise of Seiberg duality, 
for theories with moduli space admitting a toric descriptions. It is 
therefore the intent of the remainder of this paper to examine and support
\begin{conjecture}
Toric duality is Seiberg duality for ${\cal N}=1$ theories with toric
moduli spaces. 
\end{conjecture}
\section{Partial Resolutions of $\IC^3/(\IZ_3 \times \IZ_3)$ and Seiberg 
duality}
Let us proceed to check more examples. So far the other known examples
of torically dual theories 
are from various partial resolutions of $\IC^3/(\IZ_3 \times \IZ_3)$. In 
particular it was found in \cite{phases} that the (complex) cones over the 
zeroth Hirzebruch surface as well as the second del Pezzo surface each 
has two toric dual pairs. We remind the reader of these theories.
\subsection{Hirzebruch Zero}
There are two torically dual theories for the cone over the zeroth
Hirzebruch surface $F_0$. The toric and quiver diagrams are given in
\fref{f:F0}, the matter content and interactions are
\beq
\label{F0}
\hspace{-1.0cm}
\ba{c|c|c}
& \mbox{Matter Content }d & \mbox{Superpotential}\\
\hline
$I$ & {\tiny \ba{c|cccccccccccc} 
	& X_1 & X_2 & X_3 & X_4 & X_5 & X_6 & X_7 & X_8 & X_9 & X_{10}
	& X_{11} & X_{12}\\ \hline
	A & -1 & 0 & -1 & 0 & -1 & 0 & 1 & 1 & -1 & 0 & 1 & 1 \\ 
	B & 0 & -1 & 0 & -1 & 1 & 0 & 0 & 0 & 1 & 0 & 0 & 0 \\
	C & 0 & 1 & 0 & 1 & 0 & 1 & -1 & -1 & 0 & 1 & -1 & -1 \\ 
	D & 1 & 0 & 1 & 0 & 0 & -1 & 0 & 0 & 0 & -1 & 0 & 0 \ea } 
	&
{\footnotesize
\ba{r}
X_{1}X_{8}X_{10}- X_{3}X_{7}X_{10}- X_{2}X_{8}X_{9}- X_{1}X_{6}X_{12}+\\ 
X_{3}X_{6}X_{11}+ X_{4}X_{7}X_{9}+ X_{2}X_{5}X_{12}- X_{4}X_{5}X_{11}
\ea
}
\\ \hline
$II$ & {\tiny \ba{c|cccccccc}
	& X_{112} & Y_{122} & Y_{222} & Y_{111} & Y_{211} & X_{121} &
X_{212} & X_{221} \\ \hline
	A & -1 & 0 & 0 & 1 & 1 & 0 & -1 & 0 \\ 
	B & 1 & -1 & -1 & 0 & 0 & 0 & 1 & 0 \\ 
	C & 0 & 0 & 0 & -1 & -1 & 1 & 0 & 1 \\
	D & 0 & 1 & 1 & 0 & 0 & -1 & 0 & -1 \ea }
& 
{\tiny
\epsilon^{ij} \epsilon^{kl}X_{i~12}Y_{k~22}X_{j~21}Y_{l~11}
}
\ea
\eeq
\EPSFIGURE[ht]{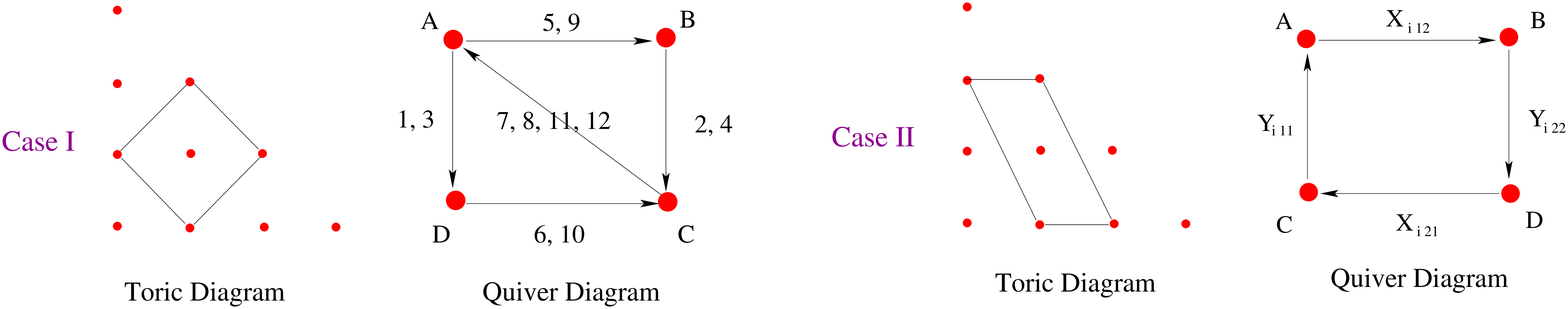,width=7.0in}
{The quiver and toric diagrams of the 2 torically dual theories
corresponding to the cone over the zeroth Hirzebruch surface $F_0$.
\label{f:F0}
}

\medskip

Let us use the field theory rules from Section 3 on Seiberg Duality to examine
these two cases in detail. The charges of the matter content for case II, 
upon promotion from $U(1)$ to $SU(N)$ 
\footnote{Concerning the $U(1)$ factors, these are in fact generically 
absent, since they are anomalous in the original $\IZ_3\times \IZ_3$ 
singularity, and the Green-Schwarz mechanism canceling their anomaly 
makes them massive \cite{iru} (see \cite{sagnan,dm,intri} for an analogous 
6d phenomenon). However, there is a well-defined sense in which one can 
use the abelian case to study the toric moduli space \cite{Morrison}.}
(for instance, following the partial resolution in the non-abelian case, 
as in \cite{Morrison,Uranga}), can be re-written as (redefining fields 
$(X_i,Y_i,Z_i,W_i) := (X_{i~12},Y_{i~22},X_{i~21},Y_{i~11})$ with $i=1,2$ 
and gauge groups $(a,b,c,d) := (A,C,B,D)$ for convenience):
$$
\ba{c|cccc}
	&	SU(N)_a & SU(N)_b  & SU(N)_c  & SU(N)_d \\
\hline
X_i	& 	\fund   & \antifund & & \\
Y_i     & 		& \fund   & \antifund & \\
Z_i     &		&	& \fund & \antifund \\
W_i	&	\antifund &	&	& \fund
\ea
$$
The superpotential is then
$$
W_{II} = X_1 Y_1 Z_2 W_2 - X_1 Y_2 Z_2 W_1 - X_2 Y_1 Z_1 W_2 + X_2 Y_2
Z_1 W_1.
$$
Let us dualise with respect to the $a$ gauge group. This is a $SU(N)$ 
theory with $N_c = N$ and $N_f = 2 N$ (as there are two $X_i$'s).
The chiral symmetry is however broken from $SU(2N)_L\times  SU(2N)_R$ to 
$SU(N)_L \times SU(N)_R$, which moreover is gauged as $SU(N)_b \times
SU(N)_d$. Ignoring the superpotential $W_{II}$, the dual theory would be:
\beq
\label{F0dualmatter}
\ba{c|cccc}
	&	SU(N)_{a'} & SU(N)_b  & SU(N)_c  & SU(N)_d \\
\hline
q_i	&	\antifund  & \fund & &\\
Y_i     &		& \fund  & \antifund & \\
Z_i	&		&	& \fund & \antifund \\
q_i'	&	\fund	& & & \antifund \\
M_{ij}	&		& \antifund & & \fund
\ea
\eeq
We note that there are $M_{ij}$ giving 4 bi-fundamentals for $bd$. They
arise from the Seiberg mesons in the bi-fundamental of the enhanced chiral 
symmetry $SU(2N) \times SU(2N)$, once decomposed with respect to the 
unbroken chiral symmetry group.
The superpotential is
$$
W' = M_{11} q_1 q_1' - M_{12} q_2 q_1' - M_{21} q_1 q_2' + 
M_{22} q_2 q_2'.
$$
The choice of signs in $W'$ will be explained shortly. 

Of course, $W_{II}$ is not zero and so give rise to a deformation in
the original theory, analogous to those studied in e.g. \cite{ahan}. In 
the dual theory, this deformation simply corresponds to $W_{II}$ rewritten in 
terms of mesons, which can be thought of as composites of the original 
quarks, i.e., 
$M_{ij} = W_i X_j$. Therefore we have 
$$
W_{II} = M_{21} Y_1 Z_2 - M_{11} 
Y_2 Z_2  - M_{22} Y_1 Z_1  + M_{12} Y_2 Z_1
$$
which is written in the new variables. The rule for the signs is that e.g. 
the field $M_{21}$ appears with positive sign in $W_{II}$, hence it 
should appear with negative sign in $W'$, and analogously for others.
Putting them together we 
get the superpotential of the dual theory
\beq
\label{F0dualsup}
\ba{ll}
W_{II}^{dual} & = W_{II} + W' = \\
& M_{11} q_1 q_1' - M_{12} q_2 q_1' - M_{21}
q_1 q_2' + M_{22} q_2 q_2' + M_{21} Y_1 Z_2 - M_{11} Y_2 Z_2  - M_{22}
Y_1 Z_1  + M_{12} Y_2 Z_1
\ea
\eeq
Upon the field redefinitions
$$
\ba{cccc}
M_{11} \rightarrow X_7 \qquad &
M_{12} \rightarrow X_8 \qquad &
M_{21} \rightarrow X_{11} \qquad & 
M_{22} \rightarrow X_{12}\\
q_1 \rightarrow X_4 \qquad &
q_2 \rightarrow X_2 \qquad &
q_{1'} \rightarrow X_9 \qquad &
q_{2'} \rightarrow X_5 \\
Y_1 \rightarrow X_6 \qquad &
Y_2 \rightarrow X_{10}\qquad &
Z_1 \rightarrow X_1\qquad &
Z_2 \rightarrow X_3
\ea
$$
we have the field content \eref{F0dualmatter} and superpotential
\eref{F0dualsup} matching precisely with case I in \eref{F0}. We
conclude therefore that the two torically dual cases I and II obtained
from partial resolutions are indeed Seiberg duals!
\subsection{del Pezzo 2}
Encouraged by the results above, let us proceed with the cone over the
second del Pezzo surface, which also have 2 torically dual theories.
The toric and quiver diagrams are given in \fref{f:dP2}.
\beq
\label{dP2}
\hspace{-1.0cm}
\ba{c|c|c}
& \mbox{Matter Content }d & \mbox{Superpotential}\\
\hline
$I$ &
	{\tiny \ba{c|ccccccccccccc}
	& Y_1 & Y_2 & Y_3 & Y_4 & Y_5 & Y_6 & Y_7 & Y_8 & Y_9 & Y_{10}
	& Y_{11} & Y_{12} & Y_{13} \\ \hline
	A & -1 & 0 & 0 & -1 & 0 & -1 & 0 & 1 & 0 & 0 & 0 & 1 & 1 \cr
	B & 0 & 0 & -1 & 0 & -1 & 1 & 0 & 0 & 0 & 1 & 0 & 0 & 0 \cr
	C & 0 & 0 & 1 & 0 & 1 & 0 & 1 & -1 & -1 & 0 & 1 & -1 & -1 \cr 
	D & 1 & -1 & 0 & 0 & 0 & 0 & 0 & 0 & 1 & -1 & 0 & 0 & 0 \cr 
	E & 0 & 1 & 0 & 1 & 0 & 0 & -1 & 0 & 0 & 0 & -1 & 0 & 0 \ea}
&
{\footnotesize
\ba{r}
Y_{2} Y_{9} Y_{11} - Y_{9} Y_{3} Y_{10} - Y_{4} Y_{8} Y_{11} -
Y_{1} Y_{2} Y_{7} Y_{13} + Y_{13} Y_{3} Y_{6} \\
- Y_{5} Y_{12} Y_{6}+
Y_{1} Y_{5} Y_{8} Y_{10} + Y_{4} Y_{7} Y_{12}
\ea
}
\\ \hline
$II$ &
	{\tiny \ba{c|cccccccccccc}
	& X_1 & X_2 & X_3 & X_4 & X_5 & X_6 & X_7 & X_8 & X_9 & X_{10}
	& X_{11} \\ \hline
	A & -1 & 0 & -1 & 0 & 0 & 0 & 1 & 0 & 0 & 0 & 1 \cr
	B & 1 & -1 & 0 & 0 & -1 & 0 & 0 & 0 & 1 & 0 & 0 \cr
	C & 0 & 0 & 1 & -1 & 0 & 1 & 0 & 0 & -1 & 0 & 0 \cr
	D & 0 & 0 & 0 & 0 & 0 & -1 & -1 & 1 & 0 & 1 & 0 \cr
	E & 0 & 1 & 0 & 1 & 1 & 0 & 0 & -1 & 0 & -1 & -1 \ea}
&
{\footnotesize
\ba{r}
X_5 X_8 X_6 X_9 + X_1 X_2 X_{10} X_7 + X_{11} X_3 X_4 \\
- X_4 X_{10} X_6 - X_2 X_8 X_7 X_3 X_9 - X_{11} X_1 X_5
\ea
}
\ea
\eeq
\EPSFIGURE[ht]{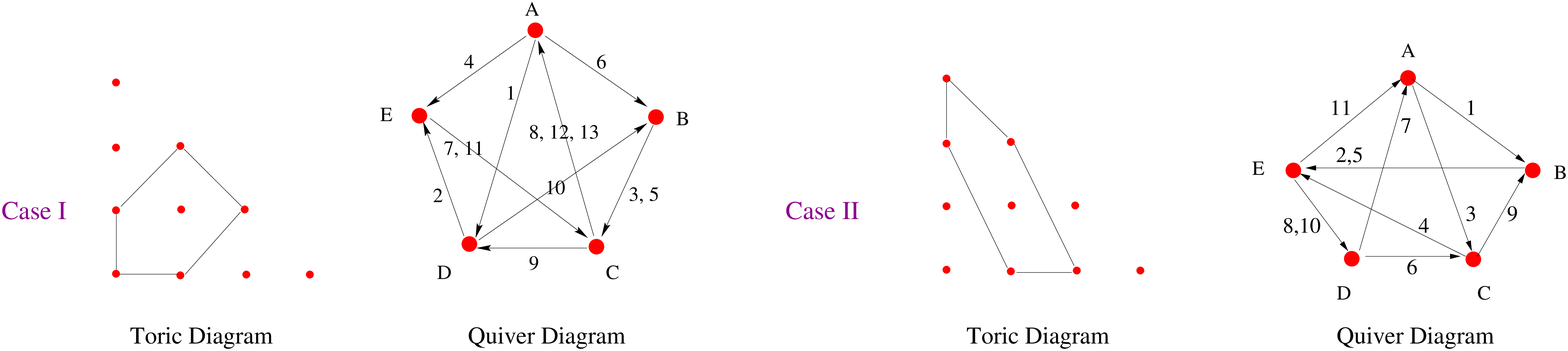,width=7in}
{The quiver and toric diagrams of the 2 torically dual theories
corresponding to the cone over the second del Pezzo surface.
\label{f:dP2}
}
Again we start with Case II. Working analogously, upon dualisation on node 
$D$ neglecting the superpotential, the matter content of II undergoes the 
following change:
\beq
\hspace{-2cm}
{\small
\label{dp2matter}
\ba{c|ccccc}
	&  SU(N)_A  & SU(N)_B & SU(N)_C & SU(N)_D & SU(N)_E \\
\hline
X_1   &  \antifund  & \fund  & & &  \\
X_2 & & \antifund & & & \fund \\
X_5 & & \antifund & & & \fund \\
X_3 & \antifund & & \fund & & \\
X_4 & & & \antifund & & \fund \\
X_9 & & \fund & \antifund & & \\
X_{11} & \fund & & & & \antifund \\
X_6 & & & \fund & \antifund &    \\
X_7 & \fund & & & \antifund &   \\
X_8 & & & & \fund & \antifund  \\
X_{10} & & & & \fund & \antifund \\
\ea
\stackrel{\mbox{dual on D}}{\Longrightarrow}
\ba{c|ccccc}
	&  SU(N)_A  & SU(N)_B & SU(N)_C & SU(N)_D & SU(N)_E \\
\hline
X_6   &  \antifund  & \fund  & & &  \\
X_5 & & \antifund & & & \fund \\
X_3 & & \antifund & & & \fund \\
X_1 & \antifund & & \fund & & \\
X_4 & & & \antifund & & \fund \\
X_{10} & & \fund & \antifund & & \\
X_{13} & \fund & & & & \antifund \\
\widetilde{X}_6 & & & \antifund & \fund &\\
\widetilde{X}_7 & \antifund & & & \fund &\\
\widetilde{X}_8 & & &&  \antifund & \fund  \\
\widetilde{X}_{10} & & &&  \antifund & \fund\\
M_{EA,1}& \fund & & & & \antifund \\
M_{EA,2}& \fund & & & & \antifund \\
M_{EC,1}&  & &\fund & & \antifund \\
M_{EC,2}&  & & \fund& & \antifund \\
\ea
}
\eeq

Let us explain the notations in \eref{dp2matter}. Before Seiberg duality we 
have 11 fields $X_{1,\ldots,11}$. After the dualisation on gauge group  
$D$, the we obtain dual quarks (corresponding to bi-fundamentals conjugate 
to the original quark $X_6,X_7, X_8,X_{10}$) which we denote 
$\widetilde{X}_6, \widetilde{X}_7, \widetilde{X}_8, \widetilde{X}_{10}$. 
Furthermore we have added meson fields $M_{EA,1},M_{EA,2},M_{EC,1},M_{EC,2}$, 
which are Seiberg mesons decomposed with respect to the unbroken chiral 
symmetry group. 

As before, one should incorporate the interactions as a deformation of 
this duality. Na\"{\i}vely we have $15$ fields in the dual theory, but
as we will show 
below, the resulting superpotential provides a mass term for the fields 
$X_4$ and $M_{EC,2}$, which transform in conjugate representations. 
Integrating them out, we will be left with 13 fields, the number of 
fields in Case I. In fact, with the mapping
$$
\ba{c|c|c|c|c|c|c|c|c|c|c|c}
\mbox{dual of II}& X_1 & X_2 & X_5 & X_3 & X_4 &X_9 & X_{11} &
\widetilde{X}_6 & \widetilde{X}_7 & \widetilde{X}_8 &
\widetilde{X}_{10} \\ \hline
\mbox{Case I} & Y_6 & Y_5 & Y_3 & Y_1 & $massive$ & Y_{10} & Y_{13} &
Y_2 & Y_4 & Y_{11} & Y_7
\ea
$$
and
%
$$
\ba{c|c|c|c|c}
\mbox{dual of II}& M_{EA,1} & M_{EA,2} & M_{EC,1} & M_{EC,2} \\
\hline
\mbox{Case I} & Y_8 & Y_{12} & Y_9 & $massive$
\ea
$$
we conclude that the matter content of the Case II dualised on gauge
group $D$ is identical to Case I!

Let us finally check the superpotentials, and also verify the claim that 
$X_4$ and $M_{EC,2}$ become massive. Rewriting the 
superpotential of II from \eref{dP2} in terms of the dual variables 
(matching the mesons as composites $M_{EA,1}=X_8 X_7$, $M_{EA,2}= X_{10} 
X_7$, $M_{EC,1}=X_8 X_6$, $M_{EC,2}= X_{10} X_6$), we have 
\begin{eqnarray*}
W_{II} & = & X_5 M_{EC,1} X_9 + X_1 X_2 M_{EA,2} + X_{11} X_3 X_4 \\
& & -X_4 M_{EC,2}- X_2 M_{EA,1} X_3 X_9 - X_{11} X_1 X_5.
\end{eqnarray*} 

As is with the previous subsection, to the above we must add the meson
interaction terms coming from Seiberg duality, namely
\begin{eqnarray*}
W_{meson} & = & M_{EA,1} \widetilde{X}_7 \widetilde{X}_8 -
M_{EA,2} \widetilde{X}_7 \widetilde{X}_{10} -M_{EC,1} \widetilde{X}_6
\widetilde{X}_8 + M_{EC,2} \widetilde{X}_6 \widetilde{X}_{10},
\end{eqnarray*}
(notice again the choice of sign in $W_{meson}$). Adding this two 
together we have
\begin{eqnarray*}
W^{dual}_{II} & = & X_5 M_{EC,1} X_9 + X_1 X_2 M_{EA,2} + X_{11} X_3 X_4 \\
& & -X_4 M_{EC,2}- X_2 M_{EA,1} X_3 X_9 - X_{11} X_1 X_5 \\
& & + M_{EA,1} \widetilde{X}_7 \widetilde{X}_8 -
M_{EA,2} \widetilde{X}_7 \widetilde{X}_{10} -M_{EC,1} \widetilde{X}_6
\widetilde{X}_8 + M_{EC,2} \widetilde{X}_6 \widetilde{X}_{10}.
\end{eqnarray*}
Now it is very clear that both $X_4$ and $M_{EC,2}$ are massive
and should be integrated out:
$$
X_4=\widetilde{X}_6 \widetilde{X}_{10},~~~~
M_{EC,2}=X_{11} X_3.
$$
Upon substitution we finally have
\begin{eqnarray*}
W^{dual}_{II} & = & X_5 M_{EC,1} X_9 + X_1 X_2 M_{EA,2} + X_{11} X_3 
\widetilde{X}_6 \widetilde{X}_{10}  - X_2 M_{EA,1} X_3 X_9 \\ 
& & - X_{11} X_1 X_5  + M_{EA,1} \widetilde{X}_7 \widetilde{X}_8 -
M_{EA,2} \widetilde{X}_7 \widetilde{X}_{10} -M_{EC,1} \widetilde{X}_6
\widetilde{X}_8,
\end{eqnarray*}
which with the replacement rules given above we obtain
\begin{eqnarray*}
W_{II}^{dual} & = & Y_3 Y_9 Y_{10} + Y_6 Y_5 Y_{12} + Y_{13} Y_1 Y_2 Y_7 
 - Y_5 Y_1 Y_{10} Y_8 \\ & & - Y_{13} Y_6 Y_3  +Y_8 Y_4 Y_{11} 
- Y_{12} Y_4 Y_7 - Y_9 Y_2 Y_{11}.
\end{eqnarray*}
This we instantly recognise, by referring to \eref{dP2}, as the
superpotential of Case I.

In conclusion therefore, with the matching of matter content and 
superpotential, the two torically dual cases I and II of the cone over
the second del Pezzo surface are also Seiberg duals.
\section{Brane Diamonds and Seiberg Duality}
Having seen the above arguments from field theory, let us support that
toric duality is Seiberg duality from yet another perspective, namely,
through brane setups. The use of this T-dual picture for D3-branes at 
singularities will turn out to be quite helpful in showing that toric 
duality reproduces Seiberg duality.

What we have learnt from the examples where a brane interval picture is
available (i.e. NS- and D4-branes in the manner of \cite{HW}) is that the 
standard Seiberg duality by brane crossing reproduces the different gauge 
theories obtained from toric arguments (different partial resolutions of a 
given singularity). Notice that the brane crossing corresponds, under 
T-duality, to a change of the $B$ field in the singularity picture, rather 
than a change in the singularity geometry \cite{conuran,Unge}. Hence, the 
two theories arise on the world-volume of D-branes probing the same 
singularity.

Unfortunately, brane intervals are rather limited, in that they can be
used to study Seiberg duality for generalized conifold singularities, 
$xy=w^kw^l$. Although this is a large class of models, not many examples 
arise in the partial resolutions of $\IC^3/(\IZ_3\times\IZ_3)$. Hence
the relation to toric duality from partial resolutions cannot be checked 
for most examples.

Therefore it would be useful to find other singularities for which a nice 
T-dual brane picture is available. Nice in the sense that there is a 
motivated proposal to realize Seiberg duality in the corresponding brane 
setup. A good candidate for such a brane setup is {\bf brane diamonds}, 
studied in \cite{aklm}. 

Reference \cite{hz} (see also \cite{hsu,hu}) introduced brane box  
configurations of intersecting NS- and NS'-branes (spanning 012345 and 
012367, respectively), with D5-branes (spanning 012346) suspended among 
them. Brane diamonds \cite{aklm} generalized (and refined) this setup by 
considering situations where the NS- and the NS'-branes recombine and span 
a smooth holomorphic curve in the $4567$ directions, in whose holes 
D5-branes can be suspended as soap bubbles. Typical brane diamond pictures 
are as in figures in the remainder of the paper. 

Brane diamonds are related by T-duality along $46$ to a large set of 
D-branes at singularities. With the set of rules to read off the 
matter content and interactions in \cite{aklm}, they provide a useful 
pictorial representation of these D-brane gauge field theories. In 
particular, they correspond to singularities obtained as the abelian 
orbifolds of the conifold studied in Section 5 of \cite{conuran}, and 
partial resolutions 
thereof. Concerning this last point, brane diamond configurations admit 
two kinds of deformations: motions of diamond walls in the directions 57, 
and motions of diamond walls in the directions 46. The former T-dualize to 
geometric sizes of the collapse cycles, hence trigger partial resolutions 
of the singularity (notice that when a diamond wall moves in 57, the 
suspended D5-branes snap back and two gauge factors recombine, leading to 
a Higgs mechanism, triggered by FI terms). The later do not  modify the 
T-dual singularity geometry, and correspond to changes in the B-fields in 
the collapsed cycles.

The last statement motivates the proposal made in \cite{aklm} for Seiberg 
duality in this setup. It corresponds to closing a diamond, while keeping 
it in the 46 plane, and reopening it with the opposite orientation. The 
orientation of a diamond determines the chiral multiplets and interactions
arising from the picture. The effect of this is shown in fig 7 of 
\cite{aklm}: The rules are 
\begin{enumerate} 
\item When the orientation of a diamond is flipped, the arrows going in 
or out of it change orientation; 
\item one has to include/remove additional arrows to ensure a good `arrow 
flow' (ultimately connected to anomalies, and to Seiberg mesons)
\item Interactions correspond to closed loops of arrows in the brane 
diamond picture.
\item In addition to these rules, and based in our experience with Seiberg 
duality, we propose that when in the final picture some mesons appear in 
gauge representations conjugate to some of the original field, the 
conjugate pair gets massive.
\end{enumerate} 
\EPSFIGURE[ht]{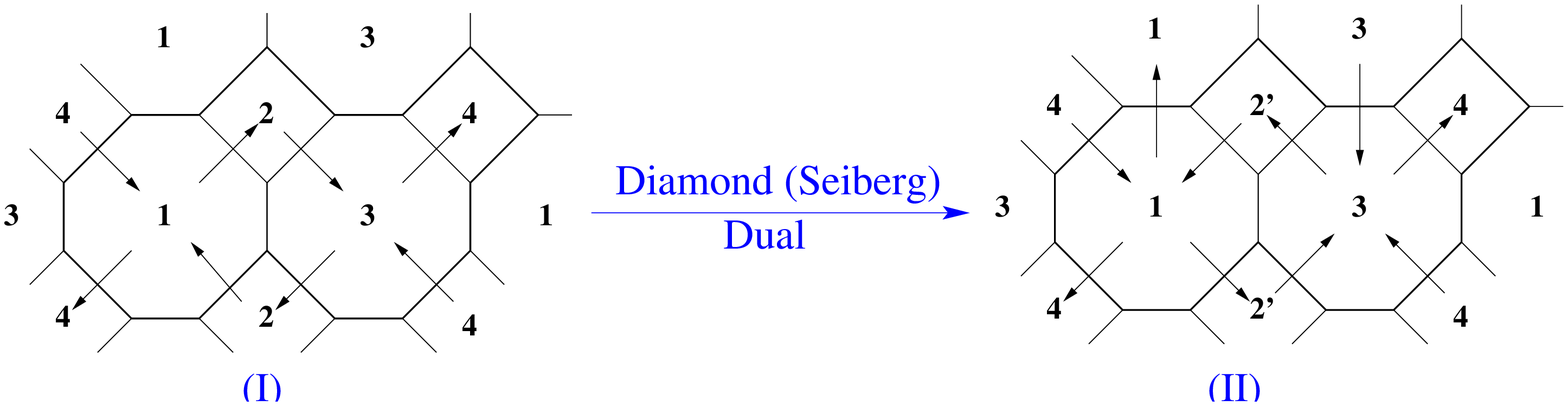,width=4in}
{Seiberg duality from the brane diamond construction for the generalized 
conifold $xy=z^2w^2$. Part (I) corresponds to the brane interval picture 
with alternating ordering of NS- and NS$'$-branes, whereas part (II) 
matches the other ordering.
\label{figz2w2}
}

These rules reproduce Seiberg duality by brane crossing in cases where a 
brane interval picture exists. In fact, one can reproduce our previous 
discussion of the $xy=z^2w^2$ in this language, as shown in figure 
\fref{figz2w2}. 
Notice that in analogy with the brane interval case the diamond transition 
proposed to reproduce Seiberg duality does not involve changes in the 
T-dual singularity geometry, hence ensuring that the two gauge theories 
will have the same moduli space. 

Let us re-examine our aforementioned examples.
\subsection{Brane diamonds for D3-branes at the cone over $F_0$}
Now let us show that diamond Seiberg duality indeed relates the two gauge 
theories arising on D3-branes at the singularity which is a complex cone 
over $F_0$.
The toric diagram of $F_0$ is similar to that of the conifold, only that 
it has an additional point (ray) in the middle of the square. Hence, it can 
be obtained from the conifold diagram by simply refining the lattice (by a 
vector $(1/2,1/2)$ if the conifold lattice is generated by $(1,0)$, 
$(0,1)$). This implies \cite{Aspinwall}) that the space can be obtained as 
a $\IZ_2$ quotient of the conifold, specifically modding $xy=zw$ by the 
action that flips all coordinates.

Performing two T-dualities in the conifold one reaches the brane diamond
picture described in \cite{aklm} (fig. 5), which is composed by two-diamond 
cell with sides identified, see Part (I) of \fref{diamcon}.
\EPSFIGURE[ht]{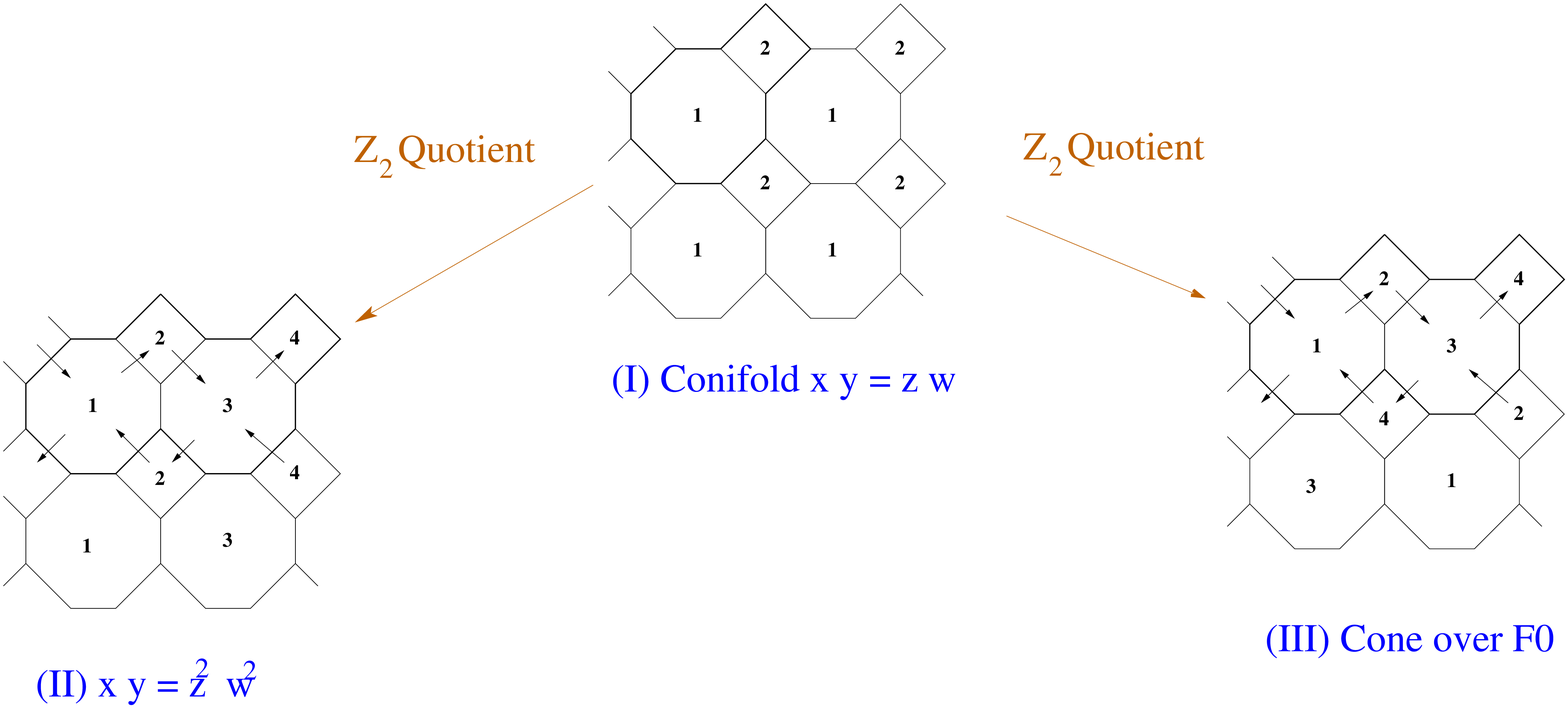,width=5in}
{
(I) Brane diamond for the conifold. Identifications in the
infinite periodic array of boxes leads to a two-diamond unit cell, whose
sides are identified in the obvious manner.
From (I) we have 2 types of $\IZ_2$ quotients: 
(II) Brane diamond for the $\IZ_2$ quotient of the conifold
$xy=z^2 w^2$, which is a case of the so-called generalised conifold. 
The identifications of sides are trivial, not tilting. The
final spectrum is the familiar non-chiral spectrum for a brane interval
with two NS and two NS' branes (in the alternate configuration);
(III) Brane diamond for the $\IZ_2$ quotient of the conifold
yielding the complex cone over $F_0$. The identifications of sides are
shifted, a fact related to the specific `tilted' refinement of the toric 
lattice.
\label{diamcon}
}
However, we are interested not in the conifold but on a $\IZ_2$ quotient 
thereof. Quotienting a singularity amounts to including more diamonds in 
the unit cell, i.e. picking a larger unit cell in the periodic array. 
There are two possible ways to do so, corresponding to two different 
$\IZ_2$ quotients of the conifold. One corresponds to the generalized 
conifold $xy=z^2w^2$ encountered above, and whose diamond picture is given 
in Part (II) of \fref{diamcon} for completeness. The second possibility
is shown in Part (III) of \fref{diamcon} and does correspond to the 
T-dual of the complex cone over $F_0$, so we shall henceforth concentrate 
on this case. Notice that the identifications of sides of the unit cell 
are shifted. The final spectrum agrees with the quiver before eq (2.2)
in \cite{toric}. Moreover, following \cite{aklm}, these fields have 
quartic interactions, associated to squares in the diamond picture, with 
signs given by the orientation of the arrow flow. They match the ones in 
case II in (\ref{F0}).

Now let us perform the diamond duality in the box labeled 2. Following the 
diamond duality rules above, we obtain the result shown in  
\fref{diamF0dual}. Careful comparison with the spectrum and interactions 
of case I in (\ref{F0}), and also with the Seiberg dual computed in 
Section 4.1 shows that the new diamond picture reproduces the toric dual / 
Seiberg dual of the initial one. Hence, brane diamond configurations 
provide a new geometric picture for this duality.

\EPSFIGURE[ht]{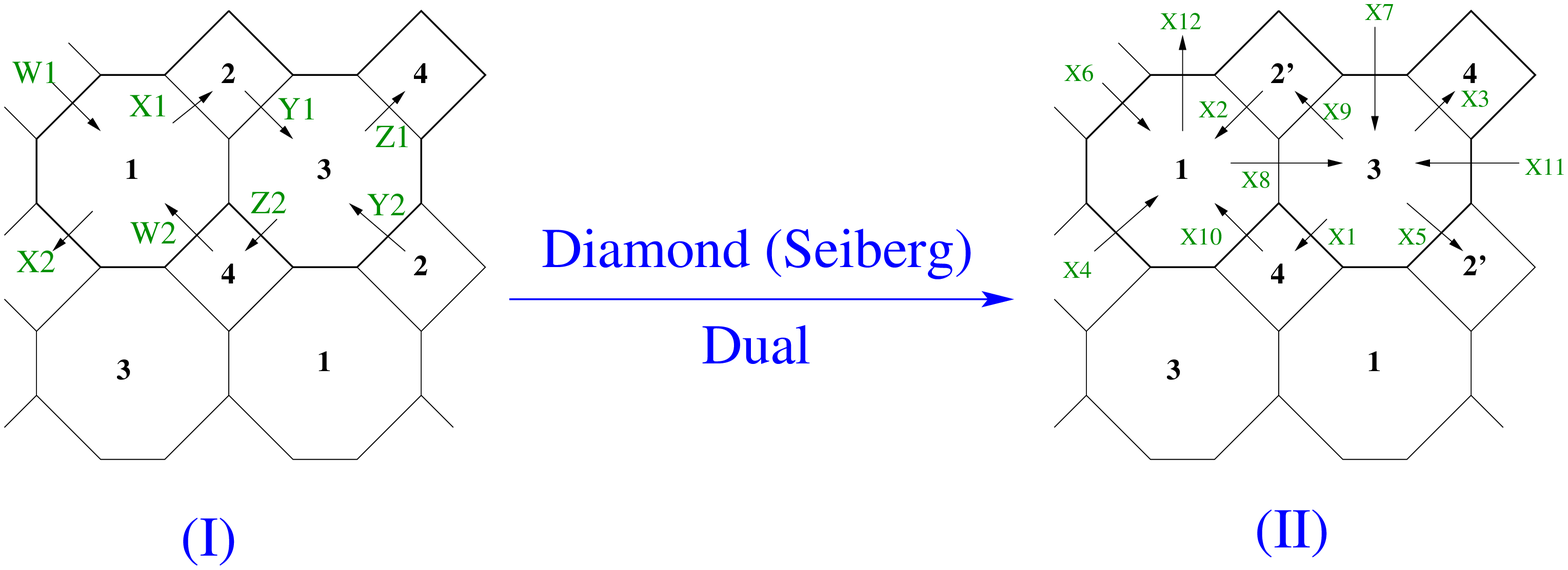,width=4in}
{Brane diamond for the two cases of the cone over $F_0$. (I) is as in 
\fref{diamcon} and (II) is the result after the diamond duality. The 
resulting spectrum and interactions are those of the toric dual (and 
also Seiberg dual) of the initial theory (I).
\label{diamF0dual}
}
\subsection{Brane diamonds for D3-branes at the cone over $dP_2$}
The toric diagram for $dP_2$ shows it cannot be constructed as a quotient 
of the conifold. However, it is a partial resolution of the orbifolded 
conifold described as $xy=v^2$, $uv=z^2$ in $\IC^5$ (we refer the
reader to \fref{condP2}.
This is a $\IZ_2\times \IZ_2$ quotient of the conifold whose brane 
diamond, shown in Part (I) of \fref{conz2z2}, contains 8 diamonds in its 
unit cell.
\EPSFIGURE[ht]{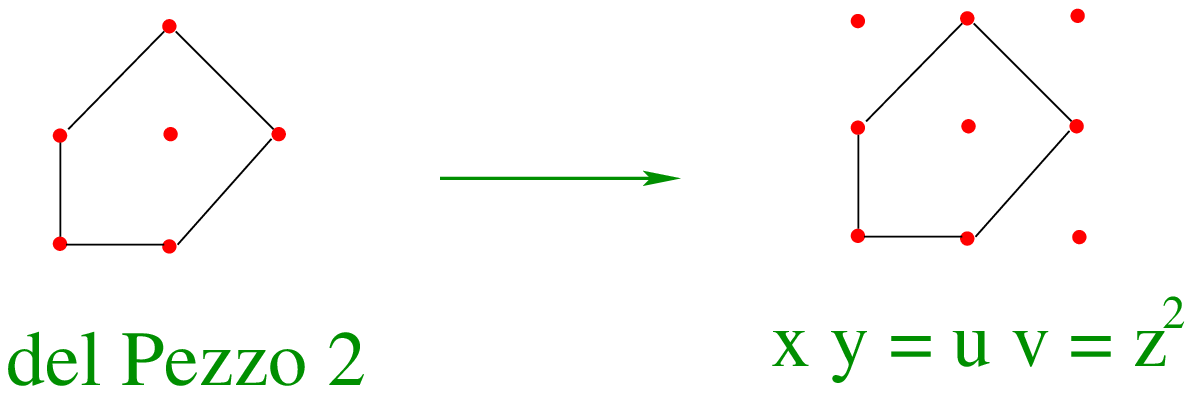,width=4in}
{Embedding the toric diagram of dP2 into the orbifolded 
conifold described as $xy=v^2$, $uv=z^2$.
\label{condP2}
}
\EPSFIGURE[ht]{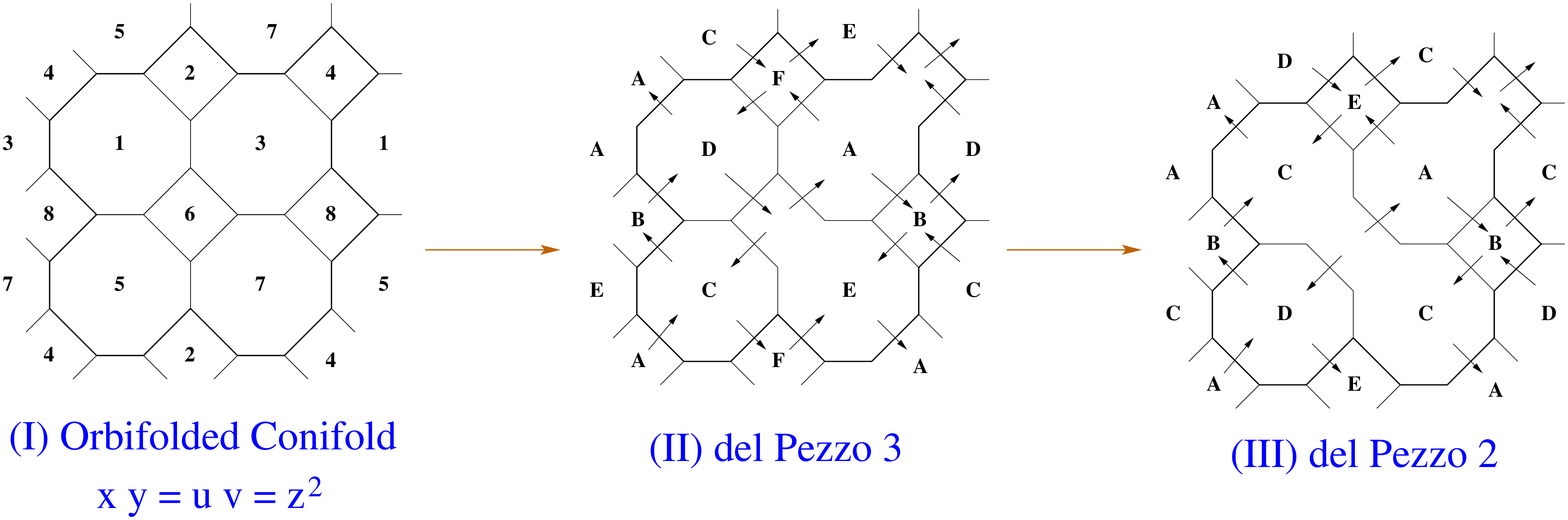,width=5.5in}
{(I) Brane diamond for a $\IZ_2\times \IZ_2$ orbifold of the
conifold, namely $xy=z^2; uv=z^2$. From this we can partial resolve to
(II) the cone over $dP3$ and thenceforth again to (III) the cone over
$dP_2$, which we shall discuss in the context of Seiberg duality.
\label{conz2z2}
}
Partial resolutions in the brane diamond language correspond to partial
Higgsing, namely recombination of certain diamonds. As usual, the 
difficult part is to identify which diamond recombination corresponds to 
which partial resolution. A systematic way proceed would 
be\footnote{As an aside, let us remark that the use of brane diamonds to 
follow partial resolutions of singularities may provide an alternative to 
the standard method of partial resolutions of orbifold singularities  
\cite{Morrison,toric}. The existence of a brane picture for partial 
resolutions of orbifolded conifolds may turn out to be a useful advantage 
in this respect.}:
\begin{enumerate}
\item Pick a diamond recombination;
\item Compute the final gauge theory;
\item Compute its moduli space, which should be the partially resolved
	singularity. 
\end{enumerate}
However, instead of being systematic, we prefer a shortcut and simply 
match the spectrum of recombined diamond pictures with known results of 
partial resolutions. In order to check we pick the right resolutions, it 
is useful to discuss the brane diamond picture for some intermediate step 
in the resolution to $dP_2$. A good intermediate point, for which the 
field theory spectrum is known is the complex cone over $dP_3$.

By trial and error matching, the diamond recombination which reproduces the 
world-volume spectrum for D3-branes at the cone over $dP_3$ (see 
\cite{toric,phases}), is shown in Part (II) of \fref{conz2z2}.
Performing a further resolution, chosen so as to match known results, one 
reaches the brane diamond picture for D3-branes on the cone over $dP_2$, 
shown in Part (III) of \fref{conz2z2}. More specifically, the spectrum and 
interactions in the brane diamond configuration agrees with those of case 
I in (\ref{dP2}).

This brane box diamond, obtained in a somewhat roundabout way, is our
starting point to discuss possible dual realizations. In fact, recall that
there is a toric dual field theory for $dP_2$, given as case II in
(\ref{dP2}). After some inspection, the desired effect is obtained by
applying diamond Seiberg duality to the diamond labeled B. The
corresponding process and the resulting diamond picture are shown in
\fref{pezzo2dual}. Two comments are in order: notice that in
applying diamond duality using the rules above, some vector-like pairs of 
fields have to be removed from the final picture; in fact one can check by 
field theory Seiberg duality that the superpotential makes them massive. 
Second, notice that in this case we are applying duality in the direction 
opposite to that followed in the field theory analysis in Section 4.2; it 
is not difficult to check that the field theory analysis works in this 
direction as well, namely the dual of the dual is the original theory. 
\EPSFIGURE{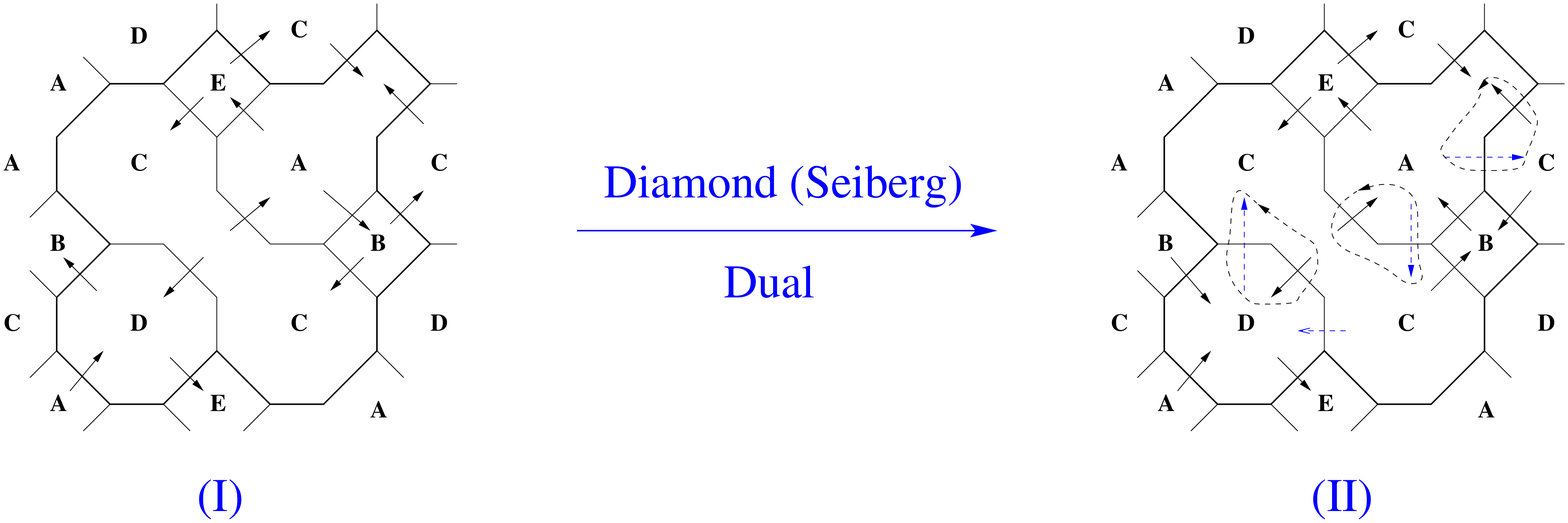,width=5in}
{The brane diamond setup for the Seiberg dual configurations of the
cone over $dP_2$. (I) is as in \fref{conz2z2} and (II) is the results
after Seiberg (diamond) duality and
gives the spectrum for the toric dual theory.
The added meson fields are drawn in dashed blue lines.
Notice that applying the diamond dual rules
carelessly one gets some additional vectorlike pairs, shown in the
picture within dotted lines. Such multiplets presumably get massive in the
Seiberg dualization, hence we do not consider them in the quiver.
\label{pezzo2dual}
}
Therefore this new example provides again a geometrical realization of 
Seiberg duality, and allows to connect it with Toric Duality.

\medskip

We conclude this Section with some remarks. The brane diamond picture 
presumably provides other Seiberg dual pairs by picking different gauge 
factors. All such models should have the same singularities as moduli 
space, and should be toric duals in a broad sense, even though all such 
toric duals may not be obtainable by partial resolutions of $\IC^3/(\IZ_3
\times \IZ_3)$. From this viewpoint we learn that Seiberg duality can 
provide us with new field theories and toric duals beyond the reach of 
present computational tools. This is further explored in Section 7.

A second comment along the same lines is that Seiberg duality on nodes for 
which $N_f\neq 2N_c$ will lead to dual theories where some gauge factors 
have different rank. Taking the theory back to the `abelian' case, some 
gauge factors turn out to be non-abelian. Hence, in these cases, even 
though Seiberg duality ensures the final theory has the same singularity 
as moduli space, the computation of the corresponding symplectic 
quotient is beyond the standard tools of toric geometry. Therefore, 
Seiberg duality can provide (`non-toric') gauge theories with toric moduli 
space.
\section{A Quiver Duality from Seiberg Duality}
If we are not too concerned with the superpotential,
when we make the 
Seiberg duality transformation, we can obtain the matter content 
very easily at the level of the quiver diagram. What we obtain are
rules for a so-called ``quiver duality'' which is a rephrasing of the
Seiberg duality transformations in field (brane diamond) theory in the 
language of quivers. Denote $(N_c)_i$ the number of colors at the $i^{th}$ 
node, and $a_{ij}$ the number of arrows from the node $i$ to the $j$ (the 
adjacency matrix)
The rules on the quiver to obtain Seiberg dual theories are
\begin{enumerate}
\item	Pick the dualisation node $i_0$. Define the following sets of 
        nodes: $I_{in} :=$ nodes having arrows going into $i_0$; 
	$I_{out} := $ those having arrow coming from $i_0$ and
	$I_{no} :=$ those unconnected with $i_0$. The node $i_0$
	should not be included in this classification. 
\item   Change the rank of the node $i_0$ from $N_c$ to $N_f-N_c$ where 
        $N_f$ is the number of vector-like flavours, $N_f=\sum\limits_{i\in 
        I_{in}} a_{i,i_0}= \sum\limits_{i\in I_{out}} a_{i_0,i}$ 
%
\item	Reverse all arrows going in or out of $i_0$, therefore
	\[
	a^{dual}_{ij} = a_{ji} \qquad \mbox{ if either }i,j = i_0
	\]
\item	Only arrows linking $I_{in}$ to $I_{out}$ will be changed and
	all others remain unaffected.
\item	For every pair of nodes $A$, $B$, $A \in I_{out}$ and  $B \in 
        I_{in}$, change the number of arrows $a_{AB}$ to
	\[
	a^{dual}_{AB} = a_{AB} - a_{i_0 A} a_{B i_0} \qquad \mbox{ for
	} A \in I_{out},~~B \in I_{in}.
	\]
	If this quantity is negative, we simply take it to mean 
        $-a^{dual}$ arrow go from $B$ to $A$. 
\end{enumerate}
These rules follow from applying Seiberg duality at the field theory 
level, and therefore are consistent with anomaly cancellation. In 
particular, notice the for any node $i\in I_{in}$, we have replaced 
$a_{i,i_0} N_c$ fundamental chiral multiplets by $-a_{i,i_0}(N_f-N_c) + 
\sum_{j\in I_{out}} a_{i,i0} a_{i_0,j}$ which equals $-a_{i,i_0}(N_f-N_c) 
+ a_{i,i0} N_f=a_{i,i_0} N_c$, and ensures anomaly cancellation in the 
final theory. Similarly for nodes $j\in I_{out}$.

It is straightforward to apply these rules to the quivers in the by now 
familiar examples in previous sections.

\medskip

In general, we can choose an arbitrary node to perform the above Seiberg 
duality rules. However, not every node is suitable for a toric description.
The reason is that, if we start from a quiver whose every node has the 
same rank $N$, after the transformation it is possible that this no longer 
holds. We of course wish so because due to the very definition of the 
$\IC^*$ action for toric varieties, toric descriptions are possible iff 
all nodes are $U(1)$, or in the non-Abelian version, $SU(N)$. If for 
instance we choose to Seiberg dualize a node with $3N$ flavours, the dual 
node will have rank $3N-N=2N$ while the others will remain with rank $N$, 
and our description would no longer be toric. For this reason we must 
choose nodes with only $2N_f$ flavors, if we are to remain within toric 
descriptions.

One natural question arises: if we Seiberg-dualise every possible allowed 
node, how many different theories will we get? Moreover how many of these
are torically dual? Let we re-analyse the examples we have thus far 
encountered.
\subsection{Hirzebruch Zero}
Starting from case $(II)$ of $F_0$ (recall \fref{F0}) all of four nodes 
are qualified to yield toric Seiberg duals (they each have 2 incoming and 2 
outgoing arrows and hence $N_f=2N$). Dualising any one will give to case 
$(I)$ of $F_0$. On the other hand, from $(I)$ of $F_0$, we see that only 
nodes $B,D$ are qualified to be dualized. Choosing either, we get back to 
the case $(II)$ of $F_0$. In another word, cases $(I)$ and $(II)$ are 
closed under the Seiberg-duality transformation. In fact, this is a very 
strong evidence that there are only two toric phases for $F_0$ no matter 
how we embed the diagram into higher $\IZ_k \times \IZ_k$ singularities. 
This also solves the old question \cite{toric,phases} that the Inverse 
Algorithm does not in principle tell us how many phases we could have. Now 
by the closeness of Seiberg-duality  transformations, we do have a way to 
calculate the number of possible phases. Notice, on the other hand, the 
existence of non-toric phases.
\subsection{del Pezzo 0,1,2}
Continuing our above calculation to del Pezzo singularities, we see that
for $dP_0$ no node is qualified, so there is only one toric phase which is 
consistent with the standard result \cite{phases} as a resolution ${\cal 
O}_{\IP^2}(-1) \rightarrow \IC^3/\IZ_3$. 
For $dP_1$, nodes $A,B$ are qualified (all notations coming from 
\cite{phases}), but the dualization gives back to same theory, so it too 
has only one phase.

For our example $dP_2$ studied earlier (recall \fref{dP2}), there are 
four points $A,B,C,D$ which are qualified in case (II). Nodes $A,C$ give 
back to case (II) while nodes $B,D$ give rise to case (I) of $dP_2$. On 
the other hand, for case (I), three nodes $B,D,E$ are qualified. Here nodes 
$B,E$ give case (II) while node $D$ give case (I). In other words, cases 
(I) and (II) are also closed under the Seiberg-duality transformation, so 
we conclude that there too are only two phases for $dP_2$, as presented 
earlier.
\subsection{The Four Phases of $dP_3$}
Things become more complex when we discuss the phases of $dP_3$. As we 
remarked before, due to the running-time limitations of the Inverse 
Algorithm, only one phase was obtained in \cite{phases}. However, one may
expect this case to have more than just one phase, and in fact a recent 
paper has given another phase \cite{HI}. Here, using the closeness 
argument we give evidence that there are four (toric) phases for $dP_3$. 
We will give only one phase in detail. Others are similarly obtained.
\EPSFIGURE[ht]{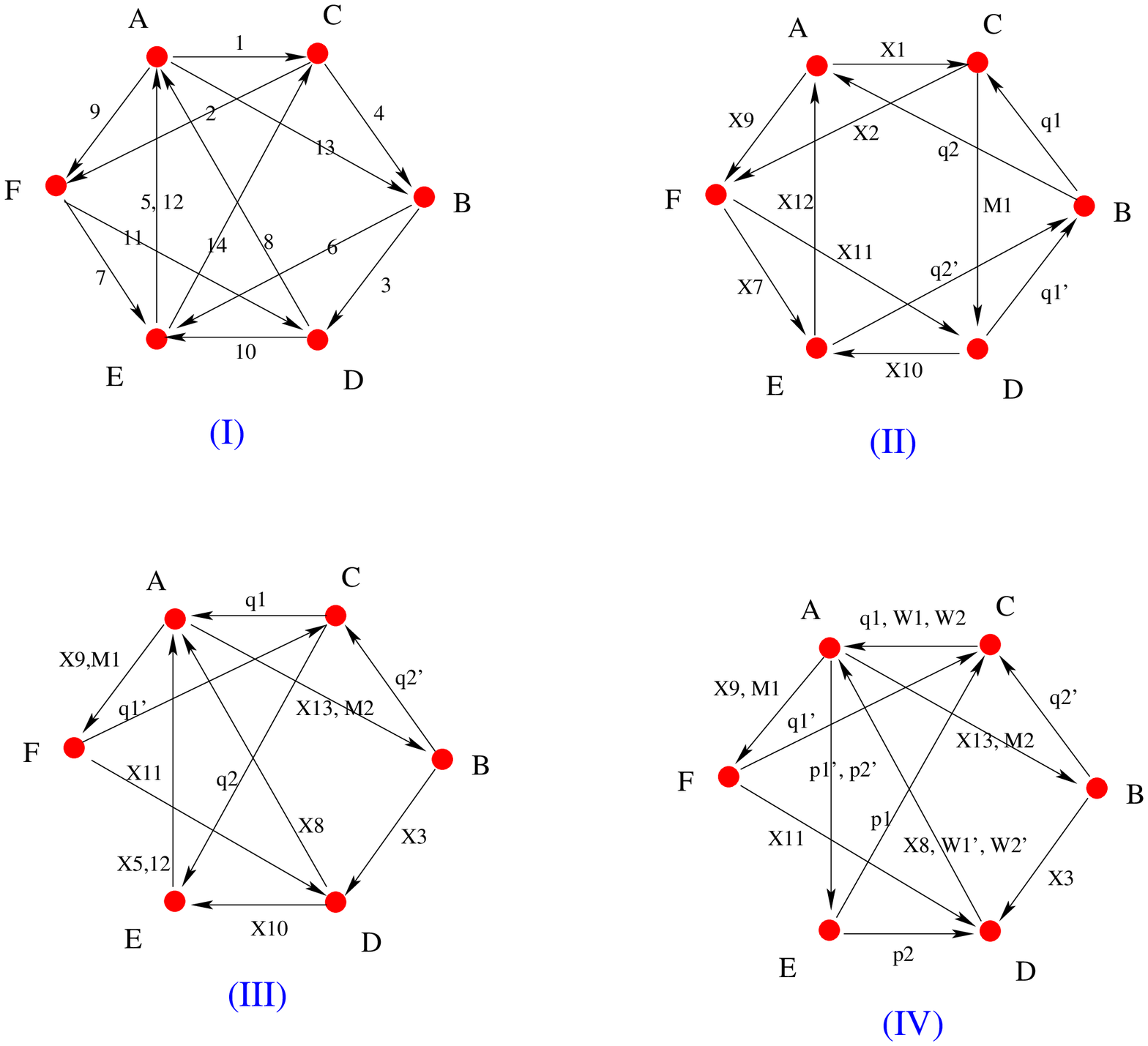,width=5in}
{The four Seiberg dual phases of the cone over $dP_3$.
\label{dP3}
}
Starting from case (I) given in \cite{phases} and dualizing node $B$,
(we refer the reader to \fref{dP3}) we get the charge (incidence) matrix $d$ as 
\[
\tmat{
 & q_1 & q_2 & q'_1 & q'_2 & X_1 & X_2 & X_7 & X_9 & X_{10} & X_{11} & 
M_1 &  X_{14}& M_2 & X_8 & M'_1 & X_5 & X_{12} &
M'_2 \cr
A & 0 & 1   & 0 & 0  & -1 & 0 & 0 & -1 & 0 & 0 & 0 & 0 & 0 & 1 & -1 &1&1
&-1  \cr
B & -1 & -1 & 1  & 1 & 0 & 0  & 0 & 0 & 0 & 0 & 0 & 0 & 0 & 0 & 0 &0&0&0
\cr
C & 1 & 0 &   0 & 0  & 1 & -1 & 0 & 0 & 0 & 0 & -1 & 1 & -1 & 0 & 0 &0 & 0
&0 \cr
D & 0 & 0   & -1 & 0 & 0 & 0 & 0 & 0 & -1 & 1 &  1 & 0  & 0 & -1 & 1 &0& 0
&0 \cr
E & 0 & 0   & 0 & -1 & 0 & 0 & 1 & 0 & 1 & 0 & 0 & -1 & 1 & 0 & 0 &-1 & -1
& 1 \cr
F & 0 & 0   & 0 & 0  & 0 & 1 & -1 & 1 &0  & -1 & 0 &  0 & 0 & 0 & 0 &0 & 0
& 0\cr
}
\]
where
$$
M_1= X_4 X_3,~~~M_2=X_4 X_6,~~~ M'_1= X_{13} X_3,~~~ M'_2= X_{13} X_6
$$
are the added mesons. Notice that $X_{14}$ and $M_2$ have opposite charge.
In fact, both are massive and will be integrate out. Same for
pairs $(X_8,M'_1)$ and $(X_5,M'_2)$.

Let us derive the superpotential. Before dual transformation, the 
superpotential is \cite{toric}
\begin{eqnarray*}
W_{I} & = & X_3 X_8 X_{13} - X_8 X_9 X_{11} - X_5 X_6 X_{13} - X_1 X_3 X_4 
X_{10} X_{12} \\
&  & X_7 X_9 X_{12} + X_4 X_6 X_{14} + X_1 X_2 X_5 X_{10} X_{11} -
X_2 X_7 X_{14}
\end{eqnarray*}
After dualization, superpotential is rewritten as
\begin{eqnarray*}
W' & = & M'_1 X_8  - X_8 X_9 X_{11} - X_5 M'_2 - X_1 M_1 
X_{10} X_{12} \\
&  & X_7 X_9 X_{12} + M_2 X_{14} + X_1 X_2 X_5 X_{10} X_{11} -
X_2 X_7 X_{14}.
\end{eqnarray*}
It is very clear that fields $X_8,M'_1,X_5,M'_2,X_{14},M_2$ are
all massive. Furthermore, we need to add the meson part
\begin{eqnarray*}
W_{meson} & = & M_1 q'_1 q_1 - M_2 q_1 q'_2 -M'_1 q'_1 q_2 + M'_2 q'_2 q_2
\end{eqnarray*}
where we determine the sign as follows: since the term $M'_1 X_8$ in 
$W'$ in positive, we need term $M'_1 q'_1 q_2$ to be negative. 
After integration all massive fields, we get the superpotential as
\begin{eqnarray*}
W_{II} & = &   -q'_1 q_2  X_9 X_{11} - X_1 M_1 X_{10} X_{12}+
X_7 X_9 X_{12}   + X_1 X_2 q'_2 q_2 X_{10} X_{11} -
X_2 X_7 q_1 q'_2 +M_1 q'_1 q_1.
\end{eqnarray*}
The charge matrix now becomes
\[
\tmat{
 & q_1 & q_2 & q'_1 & q'_2 & X_1 & X_2 & X_7 & X_9 & X_{10} & X_{11} & 
M_1 &   X_{12}  \cr
A & 0 & 1   & 0 & 0  & -1 & 0 & 0 & -1 & 0 & 0 & 0 & 1 \cr
B & -1 & -1 & 1  & 1 & 0 & 0  & 0 & 0 & 0 & 0 & 0 &  0\cr
C & 1 & 0 &   0 & 0  & 1 & -1 & 0 & 0 & 0 & 0 & -1 & 0 \cr
D & 0 & 0   & -1 & 0 & 0 & 0 & 0 & 0 & -1 & 1 &  1 & 0 \cr
E & 0 & 0   & 0 & -1 & 0 & 0 & 1 & 0 & 1 & 0 & 0 &  -1\cr
F & 0 & 0   & 0 & 0  & 0 & 1 & -1 & 1 &0  & -1 & 0 & 0 \cr
}
\]
This is in precise agreement with \cite{HI}; very re-assuring indeed!

Without further ado let us present the remaining cases.
The charge matrix for the third one (dualising node $C$ of (I)) is
\[
\tmat{
 & q_1 & q'_1 & q'_2 & q_2 & X_5 & X_{12} & X_3 & X_8 & X_9 & M_1 & X_{10}
& 
X_{11} & X_{13} & M_2 \cr
A & 1 & 0 & 0 & 0 & 1 & 1 & 0 & 1 & -1 & -1 & 0 & 0 & -1 & -1 \cr
B & 0 & 0 & -1 & 0 & 0 & 0 & -1 & 0 & 0 & 0 & 0 & 0 & 1 & 1 \cr
C & -1 & 1 & 1 & -1 & 0 & 0 & 0 & 0 & 0 & 0 & 0 & 0 & 0 & 0 \cr
D & 0 & 0 & 0 & 0 & 0 & 0 & 1 & -1 & 0 & 0 & -1 & 1 & 0 & 0 \cr
E & 0 & 0 & 0 & 1 & -1 & -1 & 0 & 0 & 0 & 0 & 1 & 0 & 0 & 0 \cr
F & 0 & -1 & 0 & 0 & 0 & 0 & 0 & 0 & 1 & 1 & 0 & -1 & 0 & 0 \cr
}
\]
with superpotential
\begin{eqnarray*}
W_{III} & = & X_3 X_8 X_{13} - X_8 X_9 X_{11} - X_5 q_2 q'_2  X_{13} 
-M_2 X_3 X_{10} X_{12} \\
& & + q_2 q'_1 X_9 x_{12} + M_1 X_5 X_{10} X_{11} -M_1 q_1 q'_1 
+M_2 q_1 q'_2.
\end{eqnarray*}

Finally the fourth case (dualising node $E$ of (III)) has the charge
matrix
\[
\tmat{
 & q_1 & W_1 & W_2 & q'_1 & q'_2 & X_3 & X_8 & W'_1 & W'_2 & X_9 & M_1 &
X_{11}& X_{13} & M_2 & p_1 & p'_1 & p'_2 & p_2 \cr
A & 1 & 1 & 1 & 0 & 0 & 0 & 1 & 1 & 1 & -1 & -1 & 0 &-1 & -1 & 0 & -1 & -1
& 0\cr
B & 0 & 0 & 0 & 0 & -1 & -1 & 0 & 0 & 0 & 0 & 0 & 0 & 1 & 1 & 0 & 0 & 0 &
0\cr
C & -1 & -1 & -1 & 1 & 1 & 0 & 0 & 0 & 0 & 0 & 0 & 0 & 0 & 0 & 1 & 0 & 0 &
0\cr
D & 0 & 0 & 0 & 0 & 0 & 1 & -1 & -1 & -1 & 0 & 0 & 1 & 0 & 0 & 0 & 0 & 0 &
1\cr
E & 0 & 0 & 0 & 0 & 0 & 0 & 0 & 0 & 0 & 0 & 0 & 0 & 0 & 0 & -1 & 1 & 1 &
-1 \cr
F & 0 & 0 & 0 & -1 & 0 & 0 & 0 & 0 & 0 & 1 & 1 & -1 & 0 & 0 & 0 & 0 & 0 &
0  \cr
}
\]
with superpotential
\begin{eqnarray*}
W_{IV} & = & X_3 X_8 X_{13} - X_8 X_9 X_{11} - W_1 q'_2 X_{13} -M_2 X_3
W'_2
+q'_1 X_9 W_2 + M_1 W'_1 X_{11} \\
& & -M_1 q_1 q'_1+M_2 q_1 q'_2+W_1 p_1 p'_1 - W_2 p_1 p'_2 -W'_1 p_2 p'_1
+W'_2 p_2 p'_2
\end{eqnarray*}
\section{Picard-Lefschetz Monodromy and Seiberg Duality}
In this section let us make some brief comments about Picard-Lefschetz
theory and Seiberg duality, a relation between which has been within
the literature \cite{VO}.
It was argued in \cite{Ito} that at least in the case of 
D3-branes placed on ADE conifolds \cite{gubser,lopez}
Seiberg duality for ${\cal N}=1$ SUSY gauge theories 
can be geometrised into Picard-Lefschetz monodromy.
Moreover in \cite{HI}
Toric Duality is interpreted as 
Picard-Lefschetz monodromy action on the 3-cycles.

On the level of brane setups, this interpretation seems to be reasonable.
Indeed, consider a brane crossing process in a brane 
interval picture. Two branes separated in $x^6$ approach, are exchanged, 
and move back. The T-dual operation on the singularity corresponds to 
choosing a collapsed cycle, decreasing its B-field to zero, and continuing 
to negative values. This last operation is basically the one generating 
Picard-Lefschetz monodromy at the level of homology classes. Similarly, 
the closing and reopening of diamonds corresponds to continuations past 
infinite coupling of the gauge theories, namely to changes in the T-dual 
B-fields in the collapsed cycles.

It is the purpose of this section to point out the observation that
while for restricted classes of theories the two phenomena are the
same, in general Seiberg duality and a na\"{\i}ve application of
Picard-Lefschetz (PL) monodromy do not
seem to coincide. We leave this issue here as a puzzle, which we shall
resolve in an upcoming work.

The organisation is as follows.
First we briefly introduce the
concept of Picard-Lefschetz monodromy for the convenience of the
reader and to establish some notation.
Then we give two examples: the first is one with two Seiberg dual
theories not related by PL and the second, PL dual theories not
related by Seiberg duality.
\subsection{Picard-Lefschetz Monodromy}
We first briefly remind
the reader of the key points of the PL theory \cite{Arnold}.
Given a singularity on a manifold $M$ and a basis $\{ \Delta_i \} \subset 
H_{n-1}(M)$ for its vanishing $(n-1)$-cycles,
going around these vanishing cycles induces 
a monodromy, acting on arbitrary cycles
$a \in H_\bullet(M)$; moreover this action is computable in terms of
intersection $a \circ \Delta_i$ of the cycle $a$ with the basis:
\begin{theorem}
\label{PL}
The monodromy group of a singularity is generated by the
Picard-Lefschetz operators $h_i$, corresponding to a 
basis $\{ \Delta_i \} \subset H_{n-1}$ of vanishing cycles.
In particular for
any cycle $a \in H_{n-1}$ (no summation in $i$)
\[
h_i(a) = a + (-1)^{\frac{n(n+1)}{2}} (a \circ \Delta_i) \Delta_i.
\]
\end{theorem}
More concretely, the PL monodromy operator
$h_i$ acts as a matrix $(h_i)_{jk}$ on the basis $\Delta_j$:
$$
h_i(\Delta_j)=(h_i)_{jk} \Delta_k.
$$

Next we establish the relationship between this geometric concept and 
a physical interpretation. 
According geometric engineering, when a
D-brane wraps a vanishing cycle in the basis, 
it give rise to a simple factor in the product gauge group. Therefore
the total number of vanishing
cycles gives the number of gauge group factors. Moreover, 
the rank of each particular factor
is determined by how many times it wraps that cycle.

For example, an original theory with gauge group $\prod\limits_j
SU(M_j)$ is represented by the brane wrapping the cycle $\sum\limits_j
M_j\Delta_j$.
Under PL monodromy, the cycle undergoes the
transformation
$$
\sum\limits_j M_j\Delta_j
\Longrightarrow \sum\limits_j M_j (h_i)_{jk} \Delta_k.
$$
Physically, the final gauge theory is 
$\prod\limits_k SU(\sum_j M_j(h_i)_{jk})$.

The above shows how the rank of the gauge theory changes under 
PL. To determine the theory completely, we also need to
see how the matter content transforms. In geometric engineering,
the matter content is given by intersection of these cycles
$\Delta_j$. Incidentally, our Inverse Algorithm gives a nice way and
alternative method of computing such intersection matrices of cycles.

Let us take $a = \Delta_j$, then
\[
h_i(\Delta_j) = \Delta_j + (\Delta_j \circ \Delta_i) \Delta_i.
\]
This is particularly useful to us because $(\Delta_j \circ \Delta_i)$,
as is well-known, is the anti-symmetrised adjacency matrix of the
quiver (for a recent discussion on this, see \cite{HI}). Indeed this
intersection matrix of (the blowup of) the vanishing homological 
cycles specifies the matter content as prescribed by D-branes wrapping
these cycles in the mirror picture. Therefore we have $(\Delta_j \circ
\Delta_i) = [a_{ji}] := a_{ji} - a_{ij}$ for $j \ne i$ and for $i=j$,
we have the self-intersection numbers $(\Delta_i \circ \Delta_i)$.
Hence we can safely write (no summation in $i$)
\beq
\label{PLmatter}
\Delta_j^{dual} = h_i(\Delta_j) = \Delta_j + [a_{ji}] \Delta_i
\eeq
for $a_{ji}$ the quiver (matter) matrix when Seiberg dualising on the
node $i$; we have also used the notation $[M]$ to mean the
antisymmetrisation $M - M^t$ of matrix $M$.
Incidentally in the basis prescribed by $\{ \Delta_i 
\}$, we have the explicit form of the Picard-Lefschetz operators in terms 
of the quiver matrix (no summation over indices):
$(h_i)_{jk} = \delta_{jk} + [a_{ji}] \delta_{ik}$.

From \eref{PLmatter} we have
\beq
\label{quiverPL}
\ba{rcl}
[a^{dual}_{jk}] & := & \Delta^{dual}_j \circ \Delta^{dual}_k =
	(\Delta_j + [a_{ji}] \Delta_i) \circ (\Delta_k + [a_{ki}]
	\Delta_i)\\
	& = & [a_{jk}] + [a_{ki}][a_{ji}] + [a_{ji}][a_{ik}] + 
		[a_{ji}][a_{ki}]\Delta_i \circ \Delta_i\\
	& = & [a_{jk}] + c_i [a_{ij}] [a_{ki}]
\ea
\eeq
where $c_i := \Delta_i \circ \Delta_i$, are constants depending only on
self-intersection.

We observe that our quiver duality rules obtained from field
theory (see beginning of Section 6) seem to resemble
\eref{quiverPL}, i.~e.~when $c_i = 1$ and $j,k \ne i$. 
However the precise relation of trying to reproduce Seiberg duality
with PL theory still remains elusive.
\subsection{Two Interesting Examples}
However the situation is not as simple. In the following we shall argue
that while Seiberg duality and a straightforward 
Picard-Lefschetz transformation
certainly do have common features and that in restricted classes of
theories such as those in \cite{Ito}, for general singularities the
two phenomena may bifurcate.

We first present two theories related by Seiberg duality that cannot
be so by Picard-Lefschetz. Consider the standard $\IC^3/\IZ_3$ theory
with $a_{ij} = \tmat{0 & 0 & 3 \cr 3 & 0 & 0 \cr 0 & 3 & 0 \cr}$ and
gauge group $U(1)^3$, given in (a) of \fref{z3}.
Let us Seiberg-dualise on node $A$ to obtain a theory (b), with
matter content $a_{ij}^{dual} = \tmat{ 0
& 3 & 0 \cr 0 & 0 & 6 \cr 3 & 0 & 0 \cr}$ and gauge group $SU(2)
\times U(1)^2$. Notice especially that the rank of the gauge 
group factors in part (b) are $(2,1,1)$ while those in  part (a) are
$(1,1,1)$. Therefore theory (b) has total rank 4 while (a) has only 3.
Since geometrically PL only shuffles the vanishing cycles and
certainly preserves their number, we see that (a) and (b) cannot be
related by PL even though they are Seiberg duals.
\EPSFIGURE{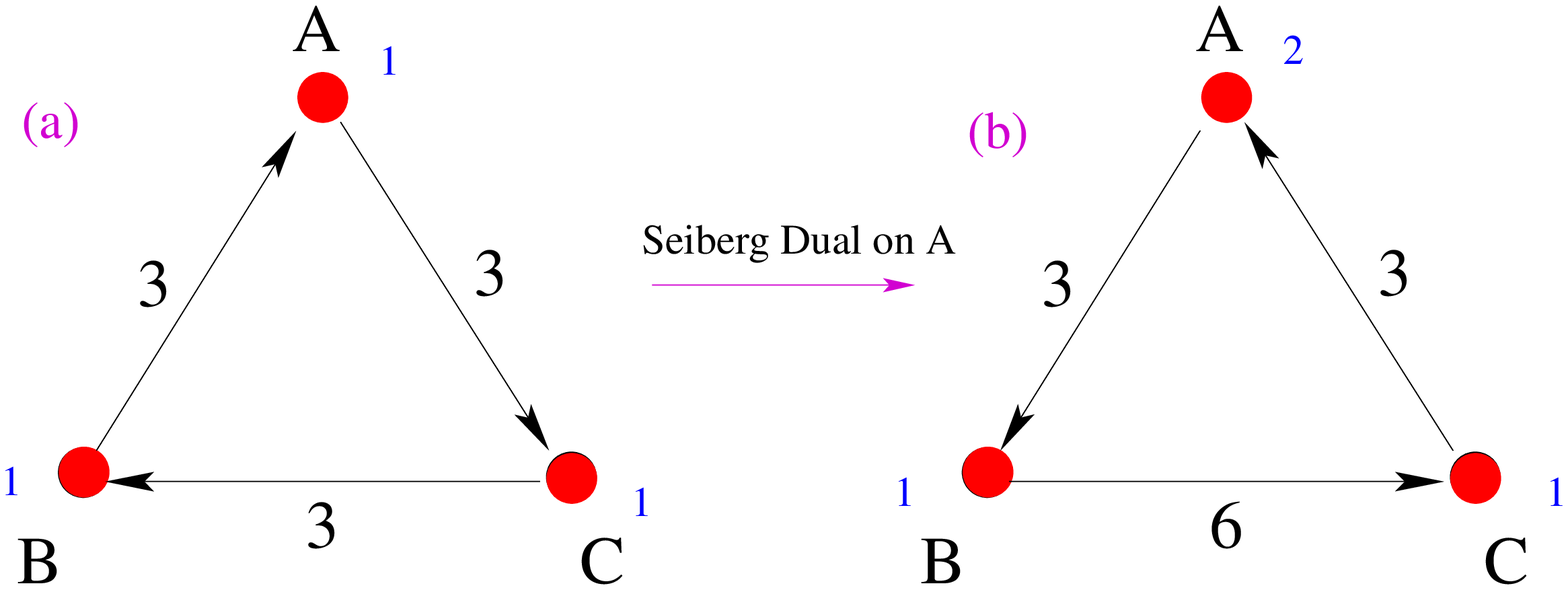,width=5in}
{Seiberg Dualisation on node $A$ of the $\IC^3/\IZ_3$ orbifold
theory. The subsequent theory cannot be obtained by a Picard-Lefschetz
monodromy transformation.
\label{z3}
}

On the other hand we give an example in the other direction, namely
two Picard-Lefschetz dual theories which are not Seiberg
duals. Consider the case given in \fref{f:NAdP3}, this is a phase of
the theory for the complex cone over dP3 as given in \cite{HI2}. This
is PL dual to any of the 4 four phases in \fref{dP3} in the previous
section by construction with $(p,q)$-webs. Note that the total rank
remains 6 under PL even though the number of nodes changed. However
Seiberg duality on any of the allowed node on any of the 4 phases
cannot change the number of nodes. Therefore, this example in
\fref{f:NAdP3} is not Seiberg dual to the other 4.

What we have learnt in this short section is that Seiberg duality and
a na\"{\i}ve application of Picard-Lefschetz monodromy seem to have
discrepancies for general singularities. The resolution of this
puzzle will be delt with in a forthcoming work.
\EPSFIGURE[ht]{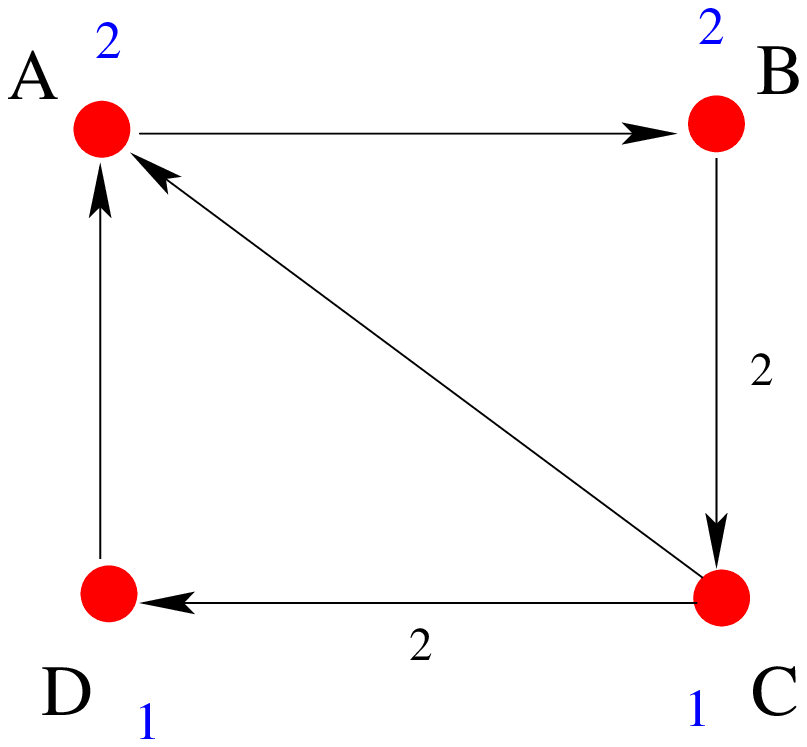,width=4.0in}
{A non-Abelian phase of the complex cone over $dP_3$. This example is
Picard-Lefschetz dual to the other 4 examples in \fref{dP3} but not
Seiberg dual thereto.
\label{f:NAdP3}
}
%
\section{Conclusions}
In \cite{toric,phases} a mysterious duality between classes of
gauge theories on D-branes probing toric singularities was
observed. Such a Toric Duality identifies the infrared moduli space of
very different theories which are candidates 
for the world-volume theory on D3-branes at threefold singularities. 
On the other hand, \cite{conuran,Unge} have recognised certain
brane-moves for brane configurations of certain toric singularities as
Seiberg duality.

In this paper we take a unified view to the above. Indeed we
have provided a physical interpretation for toric duality.
The fact that the gauge theories share by definition the same moduli space 
motivates the proposal that they are indeed physically equivalent in the 
infrared. In fact, we have shown in detail that toric dual gauge theories 
are connected by Seiberg duality.

This task has been facilitated by the use of T-dual configurations of NS 
and D-branes, in particular brane intervals and brane diamonds
\cite{aklm}. These 
constructions show that the Seiberg duality corresponds in the singularity 
picture to a change of B-fields in the collapsed cycles. Hence, the 
specific gauge theory arising on D3-branes at a given singularity, depends 
not only on the geometry of the singularity, but also on the B-field data. 
Seiberg duality and brane diamonds provide us with the tools to move 
around this more difficult piece of the singular moduli space, and 
probe different phases.

This viewpoint is nicely connected with that in \cite{toric,phases}, 
where toric duals were obtained as different partial resolutions of a 
given orbifold singularity, $\IC^3/(\IZ_3\times \IZ_3)$, leading to 
equivalent geometries (with toric diagrams equivalent up to unimodular 
transformations). Specifically, the original orbifold singularity has a 
specific assignments of B-fields on its collapsed cycles. Different 
partial resolutions amount to choosing a subset of such cycles, and 
blowing up the rest. Hence, in general different partial resolutions 
leading to the same geometric singularity end up with different 
assignments of B-fields. This explains why different gauge theories, 
related by Seiberg duality, arise by different partial resolutions.

In particular we have examined in detail the toric dual theories for
the generalised conifold $xy = z^2w^2$, the partial resolutions of
$\IC^3/(\IZ_3\times \IZ_3)$ exemplified by the complex cones over the
zeroth Hirzebruch surface as well as the second del Pezzo surface.
We have shown how these theories are equivalent under the above scheme
by explicitly having 
\begin{enumerate}
\item unimodularly equivalent toric data;
\item the matter content and superpotential related by Seiberg duality;
\item the T-dual brane setups related by brane-crossing and diamond
	duality.
\end{enumerate}
The point d'appui of this work is to show that the above three phenomena
are the same.

As a nice bonus, the physical understanding of toric duality 
has allowed us to construct 
new toric duals in cases where the partial resolution technique provided 
only one phase. Indeed the exponential running-time of the Inverse
Algorithm currently prohibits larger embeddings and partial
resolutions. Our new perspective greatly facilitates the calculation
of new phases. As an example we have constructed three new phases for
the cone over del Pezzo three one of which is in reassuring agreement
with a recent work \cite{HI} obtained from completely different methods.

Another important direction is to understand the physical meaning of 
Picard-Lefschetz transformations. 
As we have pointed out in Section 7, PL transformation and
Seiberg duality are really two different concepts even though they
coincide for certain restricted classes of theories. 
We have provided examples of two theories which are related by one but
not the other. Indeed we must pause to question ourselves. For those
which are Seiberg dual but not PL related, what geometrical action
does correspond to the field theory transformation. On the other 
hand, perhaps more importantly, for those
related to each other by PL transformation but not by
Seiberg duality, what kind of duality is realized in the 
dynamics of field theory? Does there exists a new kind of dynamical
duality not yet uncovered??

\medskip

\section*{Acknowledgements}
We would like to extend our sincere gratitude to Amer Iqbal for
discussions. 
Moreover, Y.-H.~H. would like to thank A.~Karch for enlightening
discussions.
A.~M.~U. thanks M.~Gonz\'alez for encouragement and kind support.

\bibliographystyle{JHEP}

\end{document}

%% file: texpref.tex
\newcommand{\eref}[1]{(\ref{#1})}

\newcommand{\fref}[1]{Figure~\ref{#1}}
\newcommand{\cref}[1]{Chapter~\ref{#1}}
\newcommand{\beq}{\begin{equation}}
\newcommand{\eeq}{\end{equation}}
\newcommand{\ba}{\begin{array}}
\newcommand{\ea}{\end{array}}
\newcommand{\bcenter}{\begin{center}}
\newcommand{\ecenter}{\end{center}}

\def\IB{\relax\hbox{$\inbar\kern-.3em{\rm B}$}}
\def\IC{\relax\hbox{$\inbar\kern-.3em{\rm C}$}}
\def\ID{\relax\hbox{$\inbar\kern-.3em{\rm D}$}}
\def\IE{\relax\hbox{$\inbar\kern-.3em{\rm E}$}}
\def\IF{\relax\hbox{$\inbar\kern-.3em{\rm F}$}}
\def\IG{\relax\hbox{$\inbar\kern-.3em{\rm G}$}}
\def\IGa{\relax\hbox{${\rm I}\kern-.18em\Gamma$}}
\def\IH{\relax{\rm I\kern-.18em H}}
\def\IK{\relax{\rm I\kern-.18em K}}
\def\IL{\relax{\rm I\kern-.18em L}}
\def\IP{\relax{\rm I\kern-.18em P}}
\def\IR{\relax{\rm I\kern-.18em R}}
\def\IZ{\relax\ifmmode\mathchoice
{\hbox{\cmss Z\kern-.4em Z}}{\hbox{\cmss Z\kern-.4em Z}}
{\lower.9pt\hbox{\cmsss Z\kern-.4em Z}}
{\lower1.2pt\hbox{\cmsss Z\kern-.4em Z}}\else{\cmss Z\kern-.4em Z}\fi}
\def\II{\relax{\rm I\kern-.18em I}}

\def\sCC{{\kern 0.27em\vrule height1.45ex width0.03em depth0em
          \kern-0.30em\rm C}}
\def\C{{\mathchoice
  {\sCC}
  {\sCC}
  {\kern 0.225em \vrule height1.05ex width0.025em depth0em \kern-0.25em \rm C}
  {\kern 0.180em \vrule height0.78ex width0.02em depth0em \kern-0.2em \rm C}
        }}
\def\sHH{{\rm I\kern-.16em{}H}}
\def\H{{\mathchoice
  {\sHH}
  {\sHH}
  {\rm I\kern-.13em{}H}
  {\rm I\kern-.13em{}H} }}
\def\sNN{{\rm I\kern-.16em{}N}}
\def\N{{\mathchoice
  {\sNN}
  {\sNN}
  {\rm I\kern-.12em{}N}
  {\rm I\kern-.10em{}N} }}
\def\sPP{{\rm I\kern-.16em{}P}}
\def\P{{\mathchoice
  {\sPP}
  {\sPP}
  {\rm I\kern-.12em{}P}
  {\rm I\kern-.10em{}P} }}
\def\sQQ{{\kern 0.27em \vrule height1.45ex width0.03em depth0em
          \kern-0.30em \rm Q}}
\def\Q{{\mathchoice
        {\sQQ}
        {\sQQ}
  {\kern 0.225em \vrule height1.05ex width0.025em depth0em \kern-0.25em \rm Q}
  {\kern 0.180em \vrule height0.78ex width0.020em depth0em \kern-0.20em \rm Q}
        }}
\def\sRR{{\rm I\kern-0.16em{}R}}
\def\R{{\mathchoice
  {\sRR}
  {\sRR}
  {\rm I\kern-0.12em{}R}
  {\rm I\kern-0.10em{}R} }}
\def\sZZ{{\rm Z\kern-0.32em{}Z}}
\def\Z{{\mathchoice
  {\sZZ}
  {\sZZ} 
  {\rm Z\kern-0.3em{}Z}     
  {\rm Z\kern-0.25em{}Z} }}  
\def\ZZZ{{\rm Z\kern-0.24em{}Z}}
\def\sII{{\rm I\kern-0.16em{}I}}
\def\I{{\mathchoice
  {\sII}
  {\sII}
  {\rm I\kern-0.12em{}I}
  {\rm I\kern-0.10em{}I} }}

\def\inbar{\,\vrule height1.5ex width.4pt depth0pt}
\font\cmss=cmss10 \font\cmsss=cmss10 at 7pt

\def\smiley{\hbox{\large$\bigcirc$\hspace{-0.80em}\raise.2ex
\hbox{$\cdot\cdot$}\kern-.61em\lower.2ex\hbox{\scriptsize$\smile$}}\ }
\def\frowny{\hbox{\large$\bigcirc$\hspace{-0.80em}\raise.2ex
\hbox{$\cdot\cdot$}\kern-.635em\lower.2ex\hbox{\scriptsize$\frown$}}\ }

\def\I{{\rlap{1} \hskip 1.6pt \hbox{1}}}

\newcommand{\mat}[1]{\left( \matrix{#1} \right)}
\newcommand{\tmat}[1]{{\scriptsize \mat{#1}}}
\makeatletter
\let\hangafter\@hangfrom
\makeatother

\newcommand{\drawsquare}[2]{\hbox{%
\rule{#2pt}{#1pt}\hskip-#2pt
\rule{#1pt}{#2pt}\hskip-#1pt
\rule[#1pt]{#1pt}{#2pt}}\rule[#1pt]{#2pt}{#2pt}\hskip-#2pt
\rule{#2pt}{#1pt}}

\newcommand{\fund}{\raisebox{-.5pt}{\drawsquare{6.5}{0.4}}}
\newcommand{\antifund}{\overline{\fund}}

\newtheorem{theorem}{\sf THEOREM}

\newtheorem{conjecture}{\sf CONJECTURE}